\documentclass[12pt]{article}
\usepackage{amsmath}
\usepackage{geometry}
\usepackage{hyperref}
\usepackage{graphicx}
\usepackage{caption}
\usepackage{subcaption}
\usepackage{authblk}
\usepackage{amsfonts}

\begin{document}
\title{Replica Trick Calculation for Entanglement Entropy of Static Black Hole Spacetimes}
\author[a]{Hadyan Luthfan Prihadi\footnote{hadyanluthfanp@s.itb.ac.id}}
\author[a,b]{Freddy Permana Zen\footnote{fpzen@fi.itb.ac.id}}
\author[c]{Seramika Ariwahjoedi\footnote{sera001@brin.go.id}}
\author[a,b]{Donny Dwiputra\footnote{donny.dwiputra@s.itb.ac.id}}
\affil[a]{Theoretical Physics Laboratory, Department of Physics, Institut Teknologi Bandung, Jl. Ganesha 10 Bandung, Indonesia.}
\affil[b]{Indonesia Center for Theoretical and Mathematical Physics (ICTMP), Institut Teknologi Bandung, Jl. Ganesha 10 Bandung,
	40132, Indonesia.}
\affil[c]{Research Center for Quantum Physics, National Research and Innovation Agency (BRIN) South Tangerang 15314, Indonesia.}
\maketitle
\abstract{We calculate the entanglement entropy between two (maximally-extended) spacetime regions of static black hole, seperated by horizon. As a first case, we consider the Schwarzschild black hole, and then we extend the calculations to the charged Reissner-Nordstr\"om and Schwarzschild-de Sitter black holes with more than one horizon. The case for static and spherically-symmetric solution to the more general $F(R)$ gravity is also considered. The calculation of the entanglement entropy is performed using the replica trick by obtaining the explicit form of the metric which corresponds to the replica spacetime for each black hole under consideration. The calculation of static and spherically-symmetric black holes result in the entanglement entropy that matches the Bekenstein-Hawking area law entropy.}
\newpage
\section{Introduction}
Entropy of a black hole has become a fast-growing subject in the last decades, starting from the discovery by Bekenstein \cite{Bekenstein1973} and later confirmed by Hawking's calculation \cite{Hawking_1975} which states that the entropy is proportional to the area of its horizon, or
\begin{equation}
S_{BH}=\frac{\text{Area(Horizon)}}{4G_N},
\end{equation}
where $G_N$ is the Newton's gravitational constant. The entropy of black holes can be obtained by many ways, including the semicassical path integral calculations \cite{Gibbons1977} and Hawking radiation as tunnelling \cite{Parikh2000}. In the standard statistical mechanical sense, entropy can be interpreted as the counts of the number of microstates. Such interpretation also appear in black hole thermodynamics \cite{Strominger1996}. Furthermore, the calculation of Cardy entropy by counting the horizon's states also reproduces the Bekenstein-Hawking area law entropy \cite{Solodukhin1999}. This calculation also extends to rotating Kerr black holes, which is called the Kerr/CFT correspondence \cite{Guica2009}, and even more general rotating black hole solutions for extreme black holes \cite{Sakti2018,Sakti2019,Sakti2020,Sakti2021} and for non-extreme black holes \cite{PhysRevD.82.024008,Sakti2020a}. \\
\indent On the other hand, there is another notion of entropy that is more familiar to the quantum information theory called entanglement entropy. This entropy calculates the amount of entanglement between two subsystems (or subregions if we are talking about quantum field theory) for an observer who only has access to one of the subsystems, thus it is positive if the two subsystems are entangled, and zero if it is not \cite{Calabrese_2004}. One of the most interesting features of entanglement entropy especially in conformal field theory is that it can be calculated by using the AdS/CFT correspondence \cite{Maldacena1997_LargeN,Witten1998AdSCFT,GUBSER1998105}. The result gives the entropy which is proportional to the area of a minimal surface in the bulk which is anchored to the boundary of the subregion in the boundary CFT \cite{RyuTakayanagi1,Ryu2006}. This is called the holographic entanglement entropy which has many applications due to its computational simplicity \cite{Nishioka2009}. Recently, the Ryu-Takayanagi minimal surface is generalized to the quantum extremal surface \cite{Engelhardt2015} and quantum extremal islands \cite{RevModPhys.93.035002} from the attempts to resolve the information paradox. Due to their similarity, i.e. since both are proportional to an area of a surface, it is interesting to study black hole entropy from an entanglement perspective.\\
\indent In principle, entanglement entropy can be obtained if we know how to diagonalize the reduced density matrix and obtain its eigenvalues. However, the calculation is cumbersome in many cases, especially in a continuum quantum field theory with infinite degrees of freedom (see, e.g. \cite{CasiniEntanglementLecture2022}).  Thus, it is way more convenient to derive the entanglement entropy by using the "replica trick" \cite{Calabrese_2004}. The trick is to first consider a R\'enyi entropy which is a one-parameter generalization to the entanglement entropy equipped with an extra R\'enyi index $q$, which is initially an integer, and recover the entanglement entropy after we analytically continue $q$ to real numbers and take the limit $q\rightarrow1$. In this way, our task reduces to calculating the partition function of a replicated manifold $\hat{\mathcal{M}}_q$ using path integral, which is simpler for QFT. The replica trick has many applications in statistical mechanics and can be validated by comparing the result with other methods such as the Markov chain Monte Carlo method (see \cite{Shinzato2018} and references therein). This trick has been used to calculate the entanglement entropy in a two-dimensional conformal field theory \cite{Calabrese_2009}, and to derive the holographic entanglement entropy formula \cite{Fursaev2006} which was later refined by \cite{LewkowyczMaldacena2013} and was extended to the calculation of a more general modular entropy for arbitrary $q$ \cite{Dong_2016}. We use the R\'enyi entropy instead of Tsallis entropy (see for example \cite{Calabrese_2004, Calabrese_2009}) because we use a gravitational semi-classical partition function which is simpler to calculate using the R\'enyi entropy.\\
\indent Calculating entanglement entropy of black hole spacetimes is of particular interest since it might broaden our insight into the relationship between spacetime (geometry) and entanglement, which has become a hot topic in the last decade \cite{Raamsdonk2010,Faulkner2014,Verlinde_2017EmergentGravity}. It also plays an important role in addressing problems about the consistency of quantum physics and black holes, which leads to discussions about the "black hole information paradox" (see \cite{Almheiri2021,penington_replica_2022} and references therein). It is widely known that black hole entropy can be interpreted as entanglement entropy. Entanglement entropy of a black hole with single horizon had been studied extensively \cite{Fursaev2008,Solodukhin2011}. However, the study about entanglement entropy of black hole spacetimes with multiple horizons, such as the charged Reissner-Nordstr\"om and Schwarzschild-de Sitter black holes, are still lacking. The calculations of entanglement entropy for various black holes have been done in \cite{Solodukhin2011} mainly using conical singularity contributions to the Ricci scalar. Nonetheless, in many cases, such as Schwarzschild, Reissner-Nordstr\"om, and even Kerr black holes, the Ricci scalar vanishes and thus gives no differences, especially in the gravitational calculation. Accordingly, we investigate the entanglement entropy of multi-horizon black holes from their metric. This will provide us with more understanding of the distinction between those multi-horizon black holes even though they might all share an equal Ricci scalar. We perform the work by noticing that in the replicated spacetime manifold, the mass of the original black hole becomes the function of the R\'enyi index $q$, which then contributes to the size of the deficit angle. This provides us with useful information regarding the shape of the replicated black hole spacetime. Furthermore, in contrast to the extremal limit calculations in \cite{Solodukhin2011}, in this work, we explain the transition between the extremal limit to the actual extremal entropy by a phase transition process. \\
\indent In this work, we use the replica trick to explicitly calculate the entanglement entropy of static and spherically-symmetric black hole spacetimes. We consider entanglement between two timelike regions in the maximally-extended black hole spacetimes which are connected by a wormhole with a radius equal to the black hole radius.  We calculate the entanglement entropy of the gravitational theory in the bulk by assuming that in the classical or low energy limit, the action of the bulk gravitational theory reduces to the Einstein-Hilbert action \cite{Gibbons1977,Hartle1983}. Therefore, we calculate quantum-gravitational path integral associated with the replica manifold $\hat{\mathcal{M}}_q$ using the Einstein-Hilbert action and saddle-point approximation. The quantum gravitational theory in the bulk that we are considering is just an approximation to the full theory. Indeed, we do not yet have an established quantum theory of gravity, even though some proposals have been made such as Loop Quantum Gravity (LQG) \cite{Rovelli1990,Ariwahjoedi2015,Ashtekar2021}. However, we may assume that we are calculating entanglement in a hypothetical quantum gravitational degrees of freedom using a density matrix in the approximate theory (see also \cite{LewkowyczMaldacena2013}). The calculation of entanglement entropy in gravitational spacetimes from this perspective is similar to the one that has been done in de Sitter spacetime \cite{Arias_2020}.\footnote{Although the calculation is slightly different, which will be explained more in Section 3}\\
\indent Using this way of calculating the entanglement entropy, we can investigate the entanglement entropy of more general static and spherically-symmetric black hole spacetimes such as the charged Reissner-Nordstr\"om black hole and Schwarzschild-de Sitter black hole. What distinguish those black holes from Schwarzschild's solution is that they have more than one (two, in those cases) horizons. Since there are multiple horizons, we may ask which area should correspond to the entanglement entropy and finally learn how those spacetimes are entangled. Calculations of the Reissner-Nordstr\"om and Schwarzschild-de Sitter cases are useful since, in principle, problems in more general black holes can be reduced to those two cases; the one which corresponds to the inner horizon or the one which corresponds to the cosmological horizon.\\
\indent At first, we focus on non-extreme black hole cases. However, we also try to give some remarks on the calculations of the extreme case (more general black holes with only one horizon). We find that there is a discrepancy between the entanglement entropy of extreme black holes obtained by the limiting procedure and at the extreme point itself, which is similar to the thermodynamical entropy of extreme black holes problem addressed in \cite{Carroll2009} but from the entanglement perspective. We study what aspects that need to be considered when calculating the entanglement entropy of black holes with multiple horizons which, to the authors' knowledge,  still have not been studied earlier.  There are, however, studies of black hole thermodynamics for multi-horizon black holes \cite{Xu2015,Li2017,He2018}. Nevertheless, the interpretation of the entropy that is used is still the standard thermodynamical interpretation. In this work, we are trying to look from the entanglement perspective.\\
\indent We then analyze the properties of the replica manifold for those general static and spherically-symmetric spacetimes and explicitly find the $q$-dependence of the black hole mass for the Reissner-Nordstr\"om and Schwarzschild-de Sitter black holes. The explicit form of the mass which depends on $q$ might be important for future studies regarding the calculation of the more general R\'enyi or modular entropy for arbitrary $q$, or even entropy fluctuations. However, as we show in this work, the calculation of the entanglement entropy using the replica trick only allows us to work on the value of $q$ near 1 and not for arbitrary q, and hence, by taking $q\rightarrow1$ limit, we exactly recover the Hawking entropy without any deviation. The method of calculating entanglement entropy for Reissner-Nordstr\"om and Schwarzschild-de Sitter black holes spacetime is also important as the foundation of the generalization to more general black holes with multiple horizons and higher curvature theories. Thus we also present briefly the calculations for an $F(R)$ gravity where $R$ is the Ricci scalar, without performing explicit calculations. Note that the calculations done in this paper are based on semiclassical approximation and hence we can only obtain the leading term of the entanglement entropy.\\
\indent The structure of this paper is as follows. In Section \ref{Subsec2.1} we briefly review how to calculate the entanglement entropy of some subregions in a quantum field theory using the replica trick. In Section \ref{Subsec2.2}, we use the replica trick to calculate the entanglement entropy of a Schwarzschild spacetime using a semiclassical approach. We calculate the Euclidean action and define the periodic boundary condition of the Euclidean time coordinate. At first, we show that the near-horizon geometry of the replica manifold is all we need in calculating the entanglement entropy. We then extend the former calculation to the one without using the near-horizon geometry in Section \ref{sec:2.3}, to predict what the full replica manifold looks like. Furthermore, we obtain the replica manifold for Schwarzschild spacetime perturbatively in $\varepsilon_q=1-1/q$ and calculate the entanglement entropy. \\
\indent In Section \ref{sec:3}, we extend the calculation to more general static and spherically-symmetric black hole spacetimes. The constraint for the blackening factor $f_q(r)$ in the replica manifold is shown in order to calculate the explicit form of the metric for the replica manifold for various black holes. We use the calculation to obtain the result for charged Reissner-Nordstr\"om black hole in Section \ref{sec:3.1} and for Schwarzschild-de Sitter black hole which has cosmological horizon in Section \ref{sec:3.2}. We also make some remarks on the calculation of entanglement entropy for the extreme charged Reissner-Nordstr\"om black hole with $Q=M$ in Section \ref{sec:3.1}, while the Nariai limit of the Schwarzschild-de Sitter black hole is not discussed in detail since the notion of entanglement there is not really well defined (we briefly explain why in Section \ref{sec:3.2}). We provide some calculations of the entanglement entropy for more general action such as the $F(R)$ gravity in Section \ref{sec:3.3} and conclude works in the conclusions and discussion section in Section \ref{sec:4} where we write down step-by-step procedure to calculate entanglement entropy of multi-horizon black holes from their metric.
\section{Entanglement Entropy of Schwarzschild Spacetime}
\label{Sec2}
\subsection{Entanglement Entropy and Replica Trick}
\label{Subsec2.1}
\begin{figure}
\centering
\includegraphics[scale=0.5]{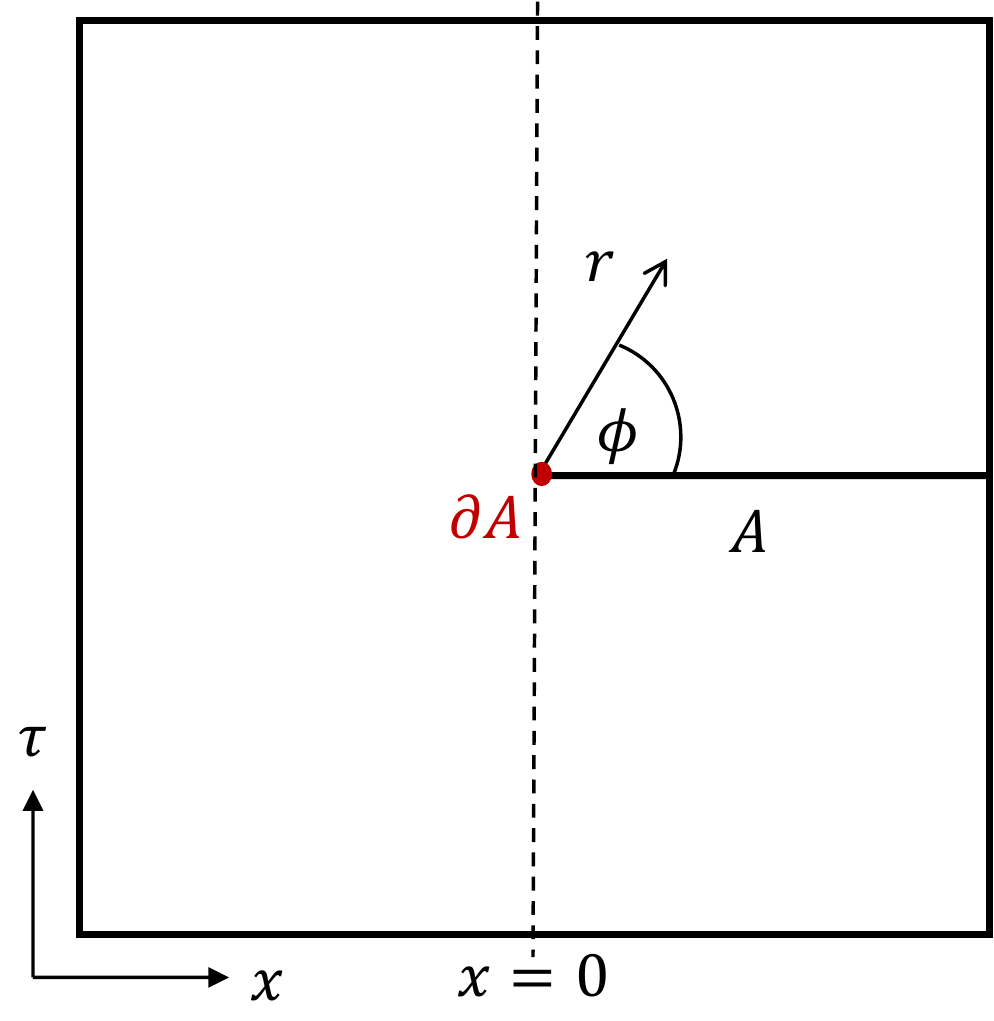}
\caption{Simple parameterization of the replica spacetime with conical singularity located at $r=0$ in flat spacetime. In this case, the region $A$ in the $\tau=0$ (Cauchy) slice is defined as $x>0$.}\label{fig:flatreplica}
\end{figure}

Entanglement entropy is defined as
\begin{equation}
S_A=-\text{Tr}(\rho_A\log\rho_A),\label{entanglemententropy}
\end{equation}
where $\rho_A$ is a reduced density matrix obtained by tracing out the degrees of freedom in the unobservable region $B$, leaving only the ones in the accessible region $A$. Calculating the entanglement entropy can be a demanding task especially for a continuous quantum field theory with infinite degrees of freedom. One of the most popular ways to calculate the entanglement entropy is to consider a one-parameter generalization of the entanglement entropy called the R\'enyi entropy, which is given by
\begin{equation}
S_{\text{R\'en}}^{(q)}=\frac{1}{1-q}\log\text{Tr}(\rho_A^q),
\end{equation}
where $q$ is an integer-valued R\'enyi index. A useful property of R\'enyi entropy is that it reduces to the entanglement entropy (\ref{entanglemententropy}) if we analytically continue $q$ into a continuous variable and take the limit $q\rightarrow1$, i.e. 
\begin{equation}
\lim_{q\rightarrow1}S_{\text{R\'en}}^{(q)}=\lim_{q\rightarrow1}\frac{1}{1-q}\log\text{Tr}(\rho_A^q)=-\partial_q \log\text{Tr}(\rho_A^q)\big|_{q=1}=S_A.\label{renyilimit}
\end{equation}
This method of calculating entanglement entropy is called replica trick (see, for example,  \cite{Calabrese_2004,Calabrese_2009,CasiniLectureEntanglementQFT}). \\
\indent Without loss of generality, suppose that the subregion $A$ is the constant-time slice of a region where $x>0$, while the complementary region, say region $B$, is the one with $x<0$ and hence the boundary $\partial A=\partial B$ is located at $x=0$. The "full" spacetime region is the manifold of the original QFT denoted as $\mathcal{M}$. The value of the trace $\text{Tr}(\rho_A^q)$ can be calculated by considering a partition function $Z[\mathcal{M}_q]$, which is represented by a quantum path integral in a replicated manifold $\mathcal{M}_q$,
which contains $q$-copy of the manifold $\mathcal{M}$ which are sewn together cyclically along the subregion $A$ following the cyclicity of the trace $\text{Tr}(\rho_A^q)$ (see, e.g., \cite{Nishioka2009}). For a normalized reduced density matrix satisfying $\text{Tr}(\rho_A)=1$, the trace $\text{Tr}(\rho_A^q)$ is then given by
\begin{equation}
\text{Tr}(\rho_A^q)=\frac{Z[\mathcal{M}_q]}{(Z[\mathcal{M}])^q},
\end{equation}
where $\mathcal{M}_1\equiv\mathcal{M}$ is nothing but the original manifold where the original QFT lives.  In the manifold $\mathcal{M}_q$, starting from the first sheet, one needs to circulate the boundary $\partial A$ $q$-times before finally arrive to the initial position. Therefore, one complete cycle of circulating $\partial A$ takes an angle of $2\pi q$ instead of $2\pi$ before arriving back to the initial position. In the CFT calculation, this corresponds to the insertion of the twist operators in $\partial A$ \cite{Calabrese_2004,Calabrese_2009}.\\
\indent We can then calculate the partition function of $\mathcal{M}_q$ by setting $\phi\sim\phi+2\pi q$ everywhere. This will introduce a conical singularity at the origin ($r=0$) with an excess angle \cite{Fursaev2006}. However, following the calculation of entropy from apparent conical singularities represented in \cite{LewkowyczMaldacena2013}, we could replace $\mathcal{M}_q$ with a single-sheeted manifold $\hat{\mathcal{M}}_q$, which has a conical singularity located at $\partial A$ with deficit angle $\Delta\phi=2\pi(1-1/q)$.  The conical singularity appears because the manifold $\hat{\mathcal{M}}_q$ is smooth if the periodicity is $\phi\sim\phi+2\pi q$, but we identify the angle with $\phi\sim\phi+2\pi$ instead. The calculations are essentially leads to the same conclusion (especially in the limit $q\rightarrow 1$) but the latter is more suitable since later on we need the information of the metric in $\hat{\mathcal{M}}_q$ for various multi-horizon black holes.\\
\indent In order to write down the metric for the manifold $\hat{\mathcal{M}}_q$, suppose that $r$ is the distance between any point around $\partial A$ to $\partial A$ and $\phi$ is the angle between $r$ and region $A$ as shown in figure (\ref{fig:flatreplica}). 
In a flat background, we could parameterize the metric of $\hat{\mathcal{M}}_q$ with polar coordinates such as
\begin{equation}
ds^2_{\hat{\mathcal{M}}_q}=\frac{r^2}{q^2}d\phi^2+dr^2+...\;,\label{metricflatq}
\end{equation}
where $...$ terms represent other transverse coordinates.  The identification of $\phi\sim\phi+2\pi$ gives $\hat{\mathcal{M}}_q$ a conical singularity located at $r=0$, i.e. at $\partial A$, with a deficit angle $\Delta\phi=2\pi(1-1/q)$ as intended. Due to the locality of the action, the relation between the partition functions $Z[\mathcal{M}_q]$ and $Z[\hat{\mathcal{M}}_q]$ is then given by
\begin{equation}
Z[\mathcal{M}_q]=\big(Z[\hat{\mathcal{M}}_q]\big)^q.
\end{equation}
From this prescription, the entanglement entropy can be calculated from
\begin{equation}
S_A=-q\partial_q\log Z[\hat{\mathcal{M}}_q]\big|_{q=1}.\label{entropyreplica}
\end{equation}
\indent There are many ways to calculate the entanglement entropy in the form of eq.  (\ref{entropyreplica}). One of them is by considering the AdS/CFT correspondence \cite{Maldacena1997_LargeN}, replacing the partition function of a CFT with a quantum gravitational partition function of the AdS spacetime by using the GKP-Witten relation \cite{GUBSER1998105,Witten1998AdSCFT}. This leads to the Ryu-Takayanagi formula of the holographic entanglement entropy \cite{RyuTakayanagi1, Ryu2006} and can be generalized to the holographic formula for R\'enyi (or modular) entropy \cite{Dong_2016}. In this work, we use eq. (\ref{entropyreplica}) to calculate the entanglement entropy in various black hole backgrounds which starts from the Schwarzschild spacetime.
\subsection{Replica Trick in Schwarzschild Spacetime}
\label{Subsec2.2}
In this section, we generalize the calculation of entanglement entropy in flat QFT into the curved Schwarzschild solution. Schwarzschild spacetime is a static and spherically symmetric solution of the vacuum Einstein equation. It describes the exterior solution of most spherically symmetric objects, including a static black hole called the Schwarzschild black hole. The Schwarzschild metric in 4 dimensions can be written as
\begin{equation}
ds^2=-\bigg(1-\frac{r_S}{r}\bigg)dt^2+\frac{dr^2}{\big(1-\frac{r_S}{r}\big)}+r^2d\Omega^2,\label{Schwarzschildmetric}
\end{equation}
where $d\Omega^2=d\theta^2+\sin^2\theta d\phi^2$ and $r_S=2G_N M$ is the Schwarzschild radius or the black hole horizon radius of a black hole with mass $M$ ($G_N$ is the usual 4-dimensional Newton's constant).  The metric (\ref{Schwarzschildmetric}) can be obtained by solving the classical equation of motion $\delta I=0$ from the Einstein-Hilbert action
\begin{equation}
I[\mathcal{M}]=\frac{1}{16\pi G_N}\int_\mathcal{M} d^4x \sqrt{|g|} R,\label{EinsteinHilbert}
\end{equation}
where $g$ is the determinant of the metric and $R$ is the Ricci scalar. Since Schwarzschild spacetime is a solution to the vacuum Einstein equation, the Ricci scalar $R$ should vanish and the contribution to the entropy should come from the Gibbons-Hawking boundary term. However, in calculating the entanglement entropy, as we will see shortly, we use the contribution from the surface term when we vary the action.
\\
\begin{figure}
\centering
\includegraphics[scale=1]{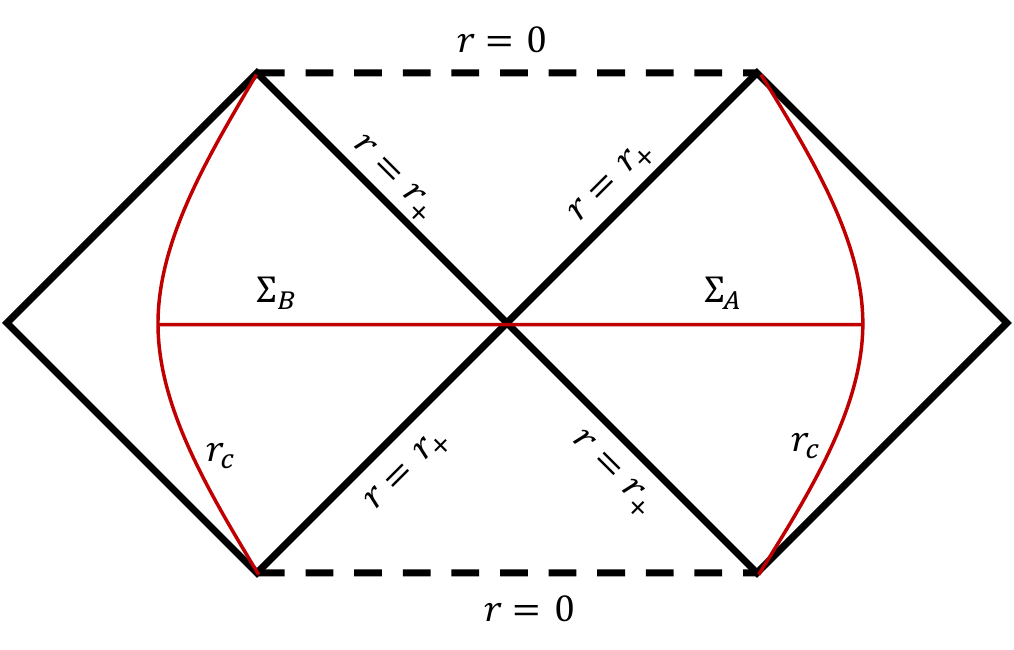}
\caption{Carter-penrose diagram of a maximally-extended Schwarzschild spacetime.}\label{fig:Schwarzschild}
\end{figure}
\indent Schwarzschild solution can be maximally-extended into two asymptotically-flat spacetimes separated by a boundary which is located at the horizon $r=r_S$. The Penrose diagram of a maximally extended Schwarzschild spacetime is depicted in figure \ref{fig:Schwarzschild}. One can thus imagine that there is a full codimension-one $t=0$ slice called $\Sigma$ which is constructed from two subregions such that $\Sigma=\Sigma_A\cup\Sigma_B$. The boundary $\partial\Sigma_A=\partial\Sigma_B$ separating $\Sigma_A$ and $\Sigma_B$ is given by a codimension-two surface $\textbf{S}^2$ with radius $r=r_S$. Nevertheless, there are other boundaries, which are located asymptotically far away from the horizon, i.e. at $r\rightarrow\infty$. However, in calculating the entanglement entropy, we only consider the boundary which separate the two subregions $A$ and $B$ for the reason that will be explained later on.\\
\indent Suppose that all points in a maximally-extended Schwarzschild spacetime $\mathcal{M}$ represent quantum gravitational degrees of freedom in the background $\mathcal{M}$. Of course, we do not yet have a full theory of quantum gravity. However,  following the action integral approach to quantum gravity \cite{Gibbons1977}, we could make an assumption by saying that in low energy or classical limit, the quantum gravitational (Euclidean) partition function is given by
\begin{equation}
Z[\mathcal{M}]=\int\mathcal{D}ge^{-I_E[g]}\approx e^{-I_E[\mathcal{M}]},\label{partitionfunction}
\end{equation}
where the second approximation comes from the saddle point approximation and $I_E[\mathcal{M}]$ is the on-shell (Euclidean) action with the metric (\ref{Schwarzschildmetric}) arise as the spherically-symmetric vacuum solution to the classical equation of motion. The Euclidean action $I_E[\mathcal{M}]$ is obtained by performing Wick rotation of the time coordinate $t\rightarrow i\tau$ of the Einstein-Hilbert action (\ref{EinsteinHilbert}). We use the Euclidean action to avoid the true singularity ($r=0$) in our calculations \cite{Gibbons1977} and also obtain the ground state of the solution \cite{Hartle1983}.\\
\begin{figure}
\centering
\includegraphics[scale=0.6]{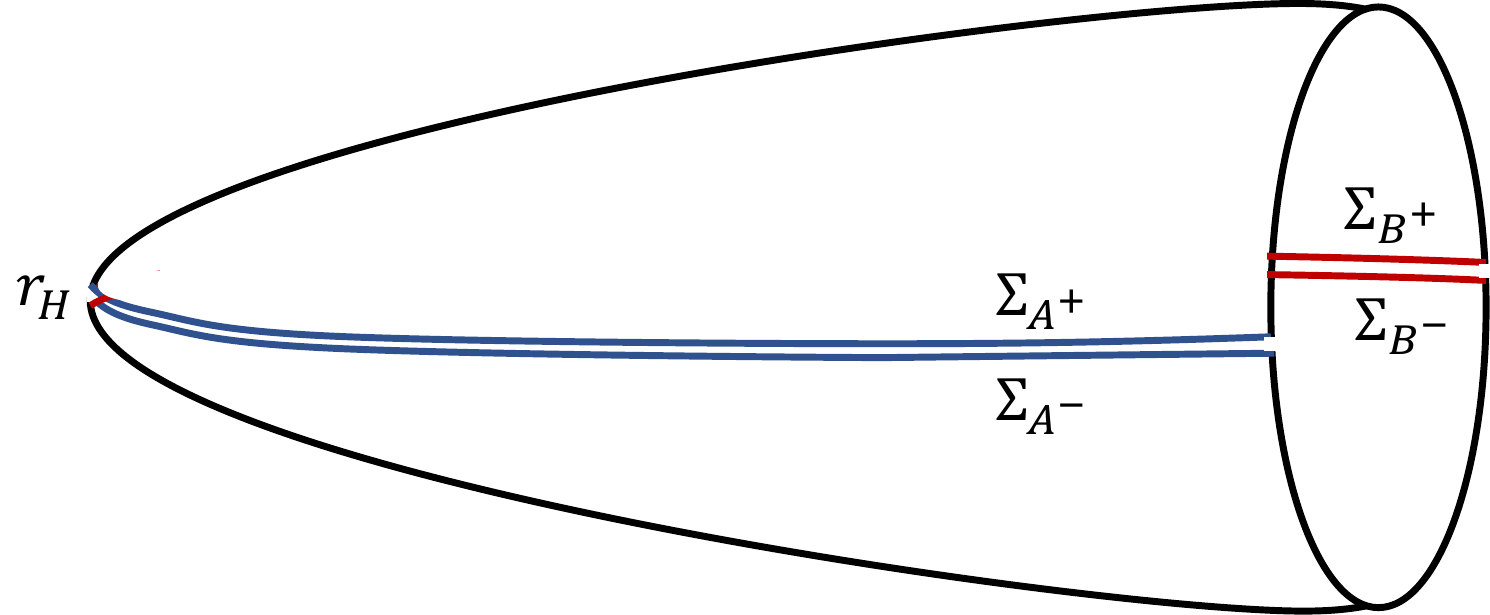}
\caption{Illustration of the total (unnormalized) density matrix $\rho$. The slice $\Sigma$ divides the cigar into two time-symmetric parts. To normalize $\rho$, we divide the calculation by $\text{Tr}(\rho)$ which is given by a full cigar.}\label{fig:rhopm}
\end{figure}
\indent In a spacetime with holographic dual such as the anti-de Sitter spacetime \cite{Maldacena1997_LargeN}, the quantum gravity partition function is more well understood since it is conjectured to be dual to a quantum partition function of a conformal field theory living in the boundary $\partial\mathcal{M}$ via the AdS/CFT correspondence \cite{Maldacena1997_LargeN, GUBSER1998105,Witten1998AdSCFT}. Other spacetimes which hypothetically have holographic duals such as de Sitter spacetime \cite{Strominger_dSCFT,Maldacena_2003,Karch_2003,Alishahiha_2004} and Friedmann-Robertson-Walker spacetime \cite{PhysRevD.80.083531} might have a better understanding of the bulk theory seen from the quantum boundary perspective. In the case that we are considering in this work (i.e. the Schwarzschild spacetime), we do not have a proper field theory description of the quantum gravitational degrees of freedom. However, we use the assumption that the quantum gravitational path integral of the theory can be described by the saddle-point approximation of the Einstein-Hilbert action in the classical limit following \cite{Gibbons1977}. Hence, we only consider entanglement of the degrees of freedom in the semiclassical approach.\\
\indent The Euclidean Schwarzschild metric is given by
\begin{equation}
ds^2=\bigg(1-\frac{r_S}{r}\bigg)d\tau^2+\frac{dr^2}{\big(1-\frac{r_S}{r}\big)}+r^2d\Omega^2.
\end{equation}
To avoid conical singularity at $r=r_S$, the Euclidean time coordinate needs to be periodic with $\tau\sim\tau+\beta$, where $\beta=8\pi M$ is the inverse temperature of the black hole. A Euclidean Schwarzschild spacetime is quite different compared to the Lorentzian counterpart. Aside from the periodic time coordinate, the Euclidean Schwarzschild spacetime is only defined for $r_S<r<\infty$ since the factor $\big(1-\frac{r_S}{r}\big)$ becomes negative in $r<r_S$, which is not allowed in a Euclidean spacetime. This can be seen from the Euclideanized Kruskal-Szekeres coordinate of the Schwarzschild spacetime which is given by
\begin{equation}
ds^2=\frac{4r_S^3}{r}e^{-r/r_S}(dT_E^2+dR^2)+r^2d\Omega^2,
\end{equation}
where the coordinates $T_E$ and $R$ obey
\begin{equation}
T_E^2+R^2=\bigg(\frac{r}{r_S}-1\bigg)e^{r/r_S}.\label{TRKruskal}
\end{equation}
The positivity of the left hand side of eq. (\ref{TRKruskal}) restricts us to the region where $r>r_S$. Following this and the periodicity of $\tau$, Euclidean Schwarzschild can be described by a so-called cigar geometry depicted in figure \ref{fig:rhopm}.\\
\begin{figure}
\begin{center}
\begin{subfigure}{.5\textwidth}
\centering 
\includegraphics[scale=0.35]{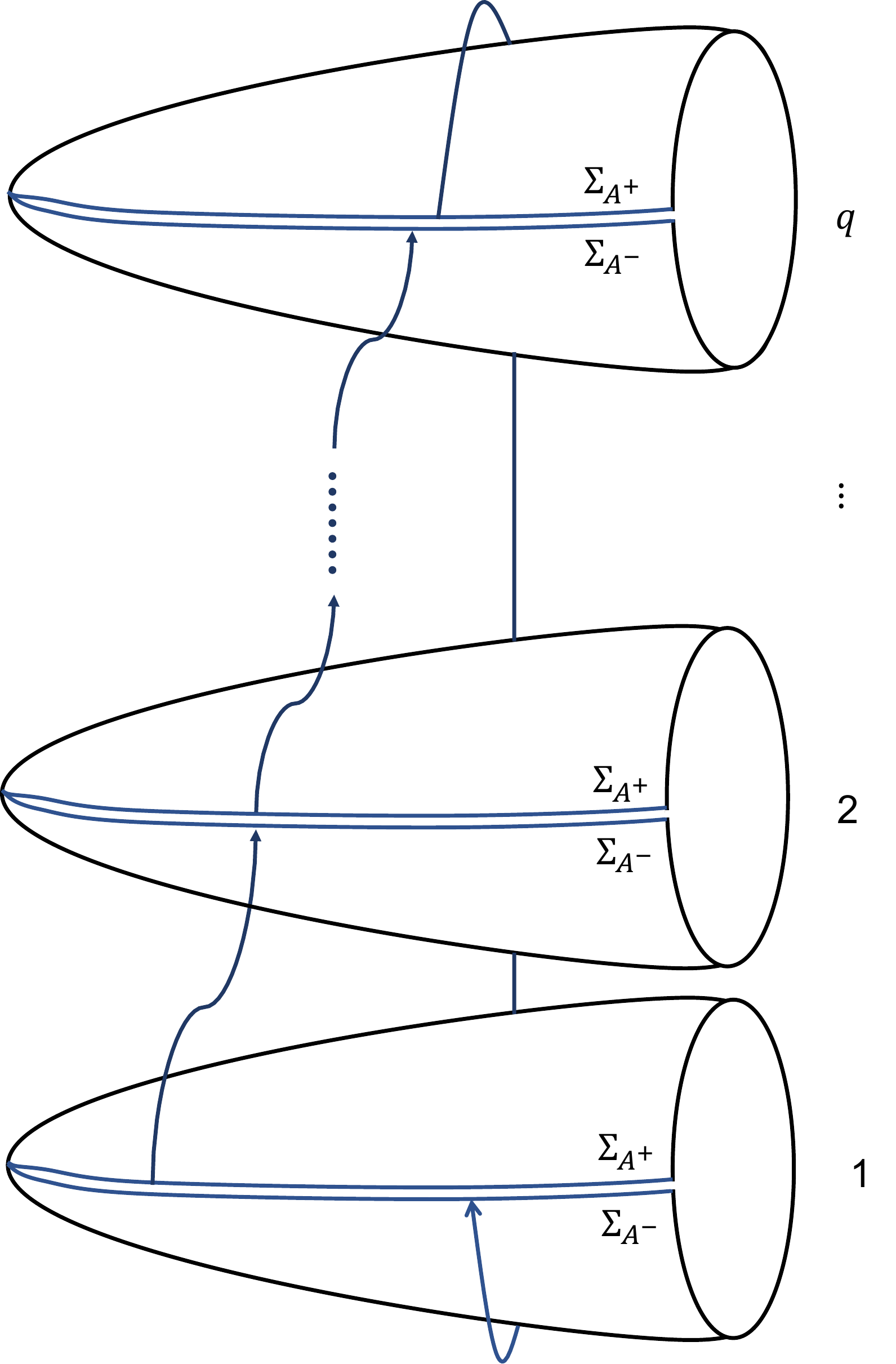}
\caption{}\label{fig:Mq}
\end{subfigure}%
\begin{subfigure}{.5\textwidth}
\centering 
\includegraphics[scale=0.35]{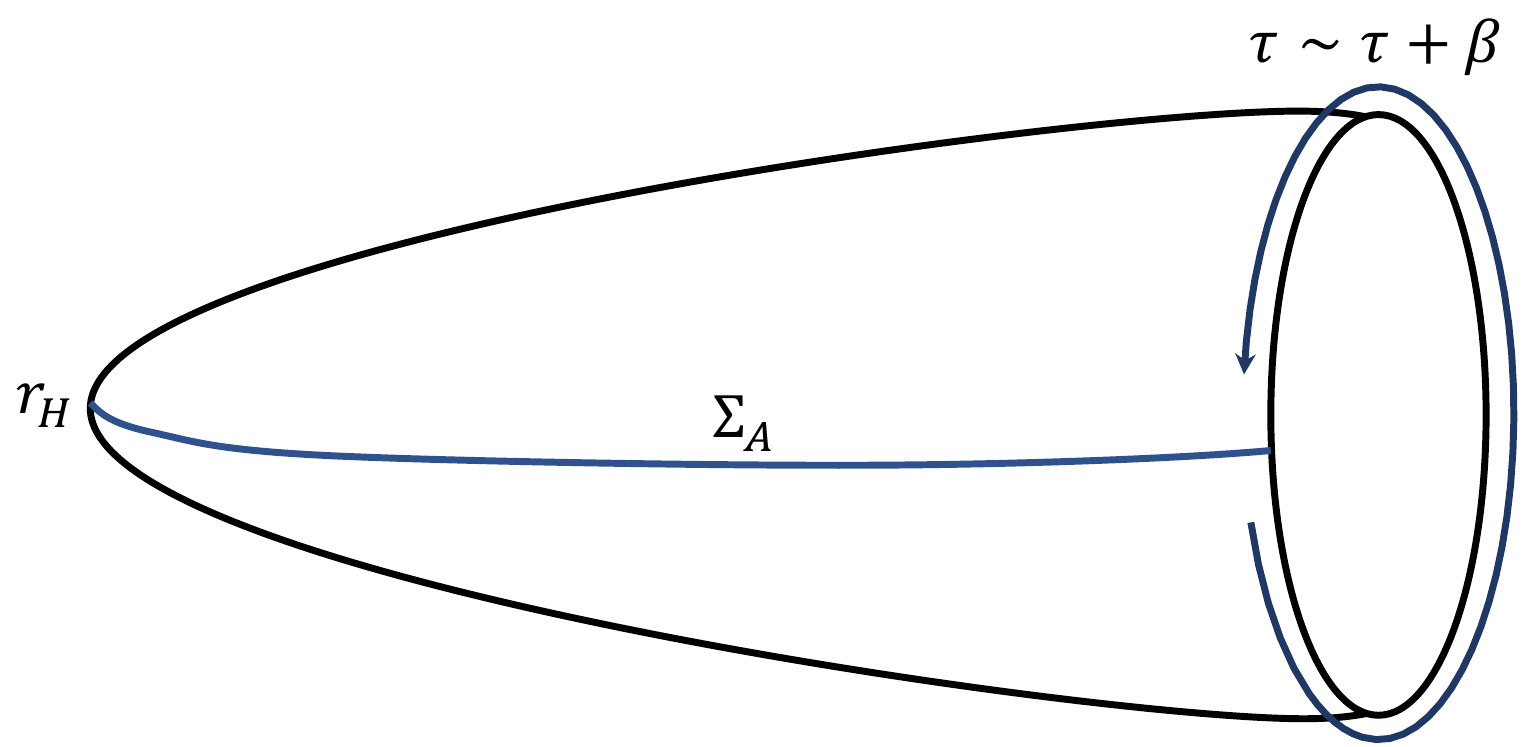}
\caption{}\label{fig:Mqhat}
\end{subfigure}
\end{center}
\caption{(a) Illustration of the partition function $Z[\mathcal{M}_q]$ of the manifold $\mathcal{M}_q$. Blue lines connected by the arrow are identified. (b) The manifold $\hat{\mathcal{M}}_q$ which now has a conical singularity at the fixed point $r=r_H$ with deficit angle $\Delta\phi=2\pi(1-1/q)$.}
\end{figure}
\indent In the cigar geometry, $\Sigma$ is a slice in the time-reflection symmetry axis that equally separates the upper and lower part, i.e. at $\tau=\tau_0,\tau_0+\beta/2$, where $\tau_0$ is arbitrary and we may set $\tau_0=0$ without loss of generality. In this case, the subregions $\Sigma_A$ and $\Sigma_B$ are the lines that start from $r=r_S$ to $r\rightarrow\infty$ at $\tau=0$ and $\tau=\beta/2$ respectively. Thus, the cigar geometry with a cut in $\Sigma$ gives us the pure (total) density matrix $\rho$ of the (hypothetical) quantum gravitational degrees of freedom in $\mathcal{M}$ (see figure \ref{fig:rhopm}). The reduced density matrix $\rho_A$ can be obtained by identifying points from $r=r_S$ to $r\rightarrow\infty$ at $\tau=\beta/2$, leaving an open cut ranging from $r=r_S$ to $r\rightarrow\infty$ at $\tau=0$. To calculate the entanglement entropy of the region $\Sigma_A$, we perform the replica trick by computing $\text{Tr}(\rho_A^q)$. The trace $\text{Tr}(\rho_A^q)$ is described by q-sheeted cigars which are cyclically-identified through the cuts following the cyclicity of the trace $\text{Tr}(\rho_A^q)$ (see figure \ref{fig:Mq}). We call this manifold $\mathcal{M}_q$ for Euclidean Schwarzschild, not to be confused with the replica manifold in flat spacetime described in the previous section.\\
\indent Following the calculation of the entanglement entropy described in Section 2.1, we replace $\mathcal{M}_q$ with $\hat{\mathcal{M}}_q$, where $\hat{\mathcal{M}}_q$ is a single manifold with conical singularity located at the origin $r=r_S$ with deficit angle $\Delta\phi=2\pi(1-1/q)$ (see figure \ref{fig:Mqhat}) and analytically-continue $q$ into non-integer value. We need to parameterize the Euclidean Schwarzschild spacetime by new coordinates which describe distance to the horizon $r=r_S$ and the angle. We will later see that, to calculate the entanglement entropy, the full metric of $\hat{\mathcal{M}}_q$ is not needed; we only need the near-horizon geometry. Suppose that $\xi\equiv r-r_S$ be the distance to the horizon while the angle is nothing but the Euclidean time $\tau$.  By replacing $r$ with $\xi$ and performing Taylor expansion near $\xi=0$, we obtain the near-horizon geometry for a Euclidean Schwarzschild,
\begin{equation}
ds^2=\frac{\xi}{r_S}d\tau^2+r_S\frac{d\xi^2}{\xi}+r_S^2d\Omega^2+...\;,
\end{equation}
where $...$ describes the subleading terms which involve higher order of $\xi$. By performing further parameterizations $y^2=\xi$, $\tilde{y}=2r_Sy$, and $\tilde{\tau}=\tau/2r_S$, we obtain
\begin{equation}
ds^2=\tilde{y}^2d\tilde{\tau}^2+d\tilde{y}^2+r_S^2d\Omega^2+...\;.\label{tildemetric}
\end{equation}
The metric near $\tilde{y}=0$ is then given by $\mathbb{R}^2\times \textbf{S}^2$. In this case, to avoid conical singularity at $\tilde{y}=0$, $\tilde{\tau}$ needs to be periodic with $\tilde{\tau}\sim\tilde{\tau}+2\pi$.\\
\indent If we want to build the metric for $\hat{\mathcal{M}}_q$, we could add the $q$-dependence to the metric in eq. (\ref{tildemetric}) such that the metric becomes similar to eq. (\ref{metricflatq}), i.e. 
\begin{equation}
ds^2=\frac{\tilde{y}^2}{q^2}d\tilde{\tau}^2+d\tilde{y}^2+r_S^2d\Omega^2+...\;,\label{tildemetricq}
\end{equation}
while keeping $\tilde{\tau}\sim\tilde{\tau}+2\pi$. In this case, we have a conical singularity located at the fixed point at $\tilde{y}=0$, with a deficit angle $\Delta\phi=2\pi(1-1/q)$. At this moment, we do not have to consider the subleading terms of the metric for the reasons that will become clear later, even if the subleading terms might contain the parameter $q$.  The metric in eq. (\ref{tildemetricq}) describes the geometry of $\hat{\mathcal{M}}_q$ near the tip of the $\hat{\mathcal{M}}_q$ cigar.\\
\indent To calculate the entanglement entropy, we use eq.  (\ref{entropyreplica}) with partition function given by eq.(\ref{partitionfunction}) for $\hat{\mathcal{M}}_q$. Therefore, we have
\begin{equation}
S_A=q\partial_qI_E[\hat{\mathcal{M}}_q]\big|_{q=1}.\label{entropypartialaction}
\end{equation}
We could obtain $\partial_qI_E[\hat{\mathcal{M}}_q]$ by first performing the variations of the action $I_E[\hat{\mathcal{M}}_q]$ with respect to the R\'enyi index $q$, and replace the variations with partial derivatives, following \cite{LewkowyczMaldacena2013}. This gives us
\begin{align}\label{derivativeaction}
\partial_qI_E[\hat{\mathcal{M}}_q]=&\int_{\hat{\mathcal{M}}_q} \frac{d^4x}{16\pi G_N}\sqrt{g}G_{\mu\nu}\partial_q g^{(q)\mu\nu}\\\nonumber&+\int_{\hat{\Omega}_\varepsilon}\frac{d^3x}{16\pi G_N}\sqrt{\gamma}\hat{n}_\rho\big(g^{(q)}_{\mu\nu}\nabla^\rho\partial_q g^{(q)\mu\nu}-\nabla_\mu\partial_q g^{(q)\rho\mu}\big),
\end{align}
where $g^{(q)}_{\mu\nu}$ is the component of the metric tensor given by eq. (\ref{tildemetricq}). The first term of the right hand side of eq. (\ref{derivativeaction}) vanishes due to the equation of motion $G_{\mu\nu}=0$, where $G_{\mu\nu}$ is the usual Einstein tensor. The second term is the surface term evaluated at a hypersurface $\hat{\Omega}_{\varepsilon}$ of constant and arbitrarily small radius $\varepsilon$ centered at $\tilde{y}=0$. This hypersurface is necessary to isolate the conical singularity from our calculations, and we take the limit of $\varepsilon\rightarrow0$ after the calculations. Here, $\gamma$ is the determinant of the induced metric in $\hat{\Omega}_\varepsilon$ and $\hat{n}^\mu$ is a unit normal vector of $\hat{\Omega}_\varepsilon$ pointing outside, away from $\tilde{y}=0$. Other surface term should appear from the boundary term at infinity. However, this term vanishes since we impose a boundary condition which states that there is no variations at the infinity.\\
\indent Directly plugging the metric in eq. (\ref{tildemetricq}) to the surface term in eq. (\ref{derivativeaction}) and integrating $\tilde{\tau}$ from $0$ to $2\pi$ gives us
\begin{equation}
\partial_qI_E[\hat{\mathcal{M}}_q]=\frac{4\pi r_S^2}{4G_Nq^2}+\mathcal{O}(\tilde{y})\bigg|_{\tilde{y}=\varepsilon}.\label{deractionresult}
\end{equation}
By taking the limit $\varepsilon\rightarrow0$, we omit all of the subleading terms contained in $\mathcal{O}(\tilde{y})$ and by eq. (\ref{entropypartialaction}), the entanglement entropy precisely recover the area law, i.e. 
\begin{equation}
S_A=\frac{(4\pi r_S^2)}{4G_N}=\frac{\text{Area(Horizon)}}{4G_N}.\label{entropyarea}
\end{equation}
The result does not depends on the Euclidean time $\tau$ and hence this is also the result for the Lorentzian spacetime. Notice that the $\mathcal{O}(\tilde{y})$ terms come from the subleading terms of the metric in eq. (\ref{tildemetricq}). Therefore, knowing the near-horizon geometry of $\hat{\mathcal{M}}_q$ is sufficient to calculate the entanglement entropy. We give some calculations on the subleading terms using the Gaussian normal coordinates in Appendix A to complete our justification. 
\\
\indent We would like to make several remarks to the result given by eq. (\ref{entropyarea}). First, this is not a surprising result; this might even rather trivial. Similar calculations of the entanglement entropy of black holes in quantum gravity has been done in \cite{Fursaev2008} and produces the same result (see also \cite{Solodukhin2011}). However, it does not perform the calculation explicitly using the metric of the replica manifold $\hat{\mathcal{M}}_q$. Instead, they directly calculate the action's variation from the Ricci scalar's delta-function contribution due to the presence of conical singularity. Nonetheless, the method of calculating the entanglement entropy using near-horizon geometry is useful to determine entanglement entropy in more general black holes, which becomes our main focus of this work. \\
\indent Second, we may also introduce an arbitrary (large but finite) cut-off $r_c$ such that the regions $\Sigma_A$ and $\Sigma_B$ ranging from $r=r_S$ to $r=r_c$ at $\tau=0$ and $\tau=\beta/2$ respectively. The addition of such cut-off does not affect the final result of the entanglement entropy since we only need the near-horizon metric (or metric near the tip of the cigar) to calculate the entanglement entropy. If the cut-off is large enough, we may then impose the boundary condition at the cut-off and take the variation of the action there to vanish. In other words, the result remains independent of the cut-off $r_c$ as long as it is large enough. This fact is also useful for calculating the entanglement entropy of black holes with more than one horizons, which will be elaborated in the next section.\\
\indent The entanglement entropy in Schwarzschild spacetime calculated in this section describes correlations which are strong near the horizon and therefore the leading term of the entanglement entropy is proportional to the area of the entangling surface $\partial\Sigma$. Any contribution from long-range correlation, i.e. entanglement between degrees of freedom which are relatively far from the entangling surface only contributes as the sub-leading term of the entanglement entropy. This physical interpretation helps us to be sure that any entangled degrees of freedom outside the cut-off region $r>r_c$ does not contributes to the entanglement entropy and the entropy is still well described by the area law of eq. (\ref{entropyarea}).
\subsection{Calculations Without Near-horizon Geometry}
\label{sec:2.3}
\begin{figure}
\begin{center}
\begin{subfigure}{.5\textwidth}
\centering 
\includegraphics[scale=0.45]{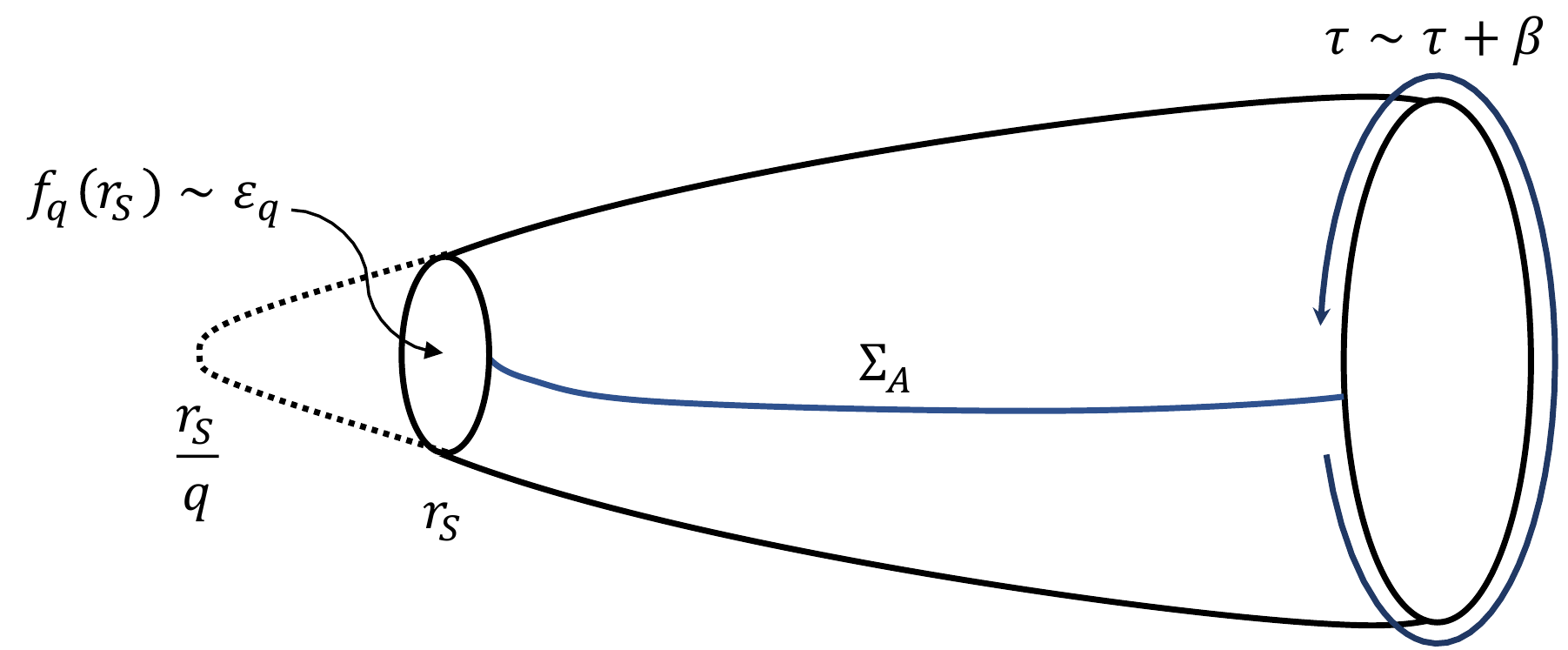}
\caption{}\label{fig:fullEuc}
\end{subfigure}%
\begin{subfigure}{.5\textwidth}
\centering 
\includegraphics[scale=0.45]{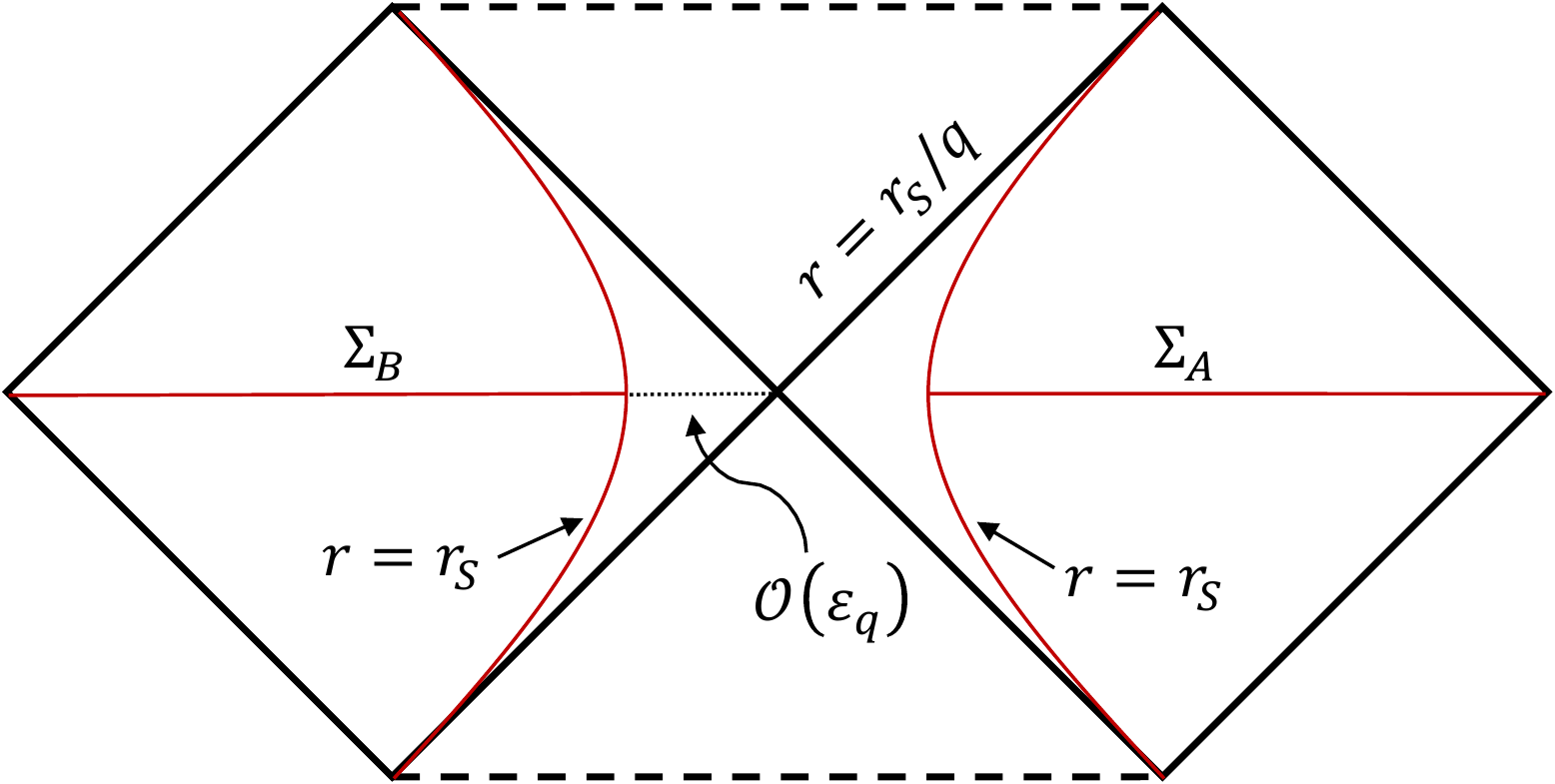}
\caption{}\label{fig:fullLor}
\end{subfigure}
\end{center}
\caption{The illustration for calculations in the manifold $\hat{\mathcal{M}}_q$. (a) represents the Euclidean calculation while (b) describes the Lorentzian Carter-Penrose diagram. In both pictures, the entangling surface $r_S$ is located slightly away from the original black hole horizon $r_S/q$, which in order $\varepsilon_q$ to be precise. In the Euclidean setting (a), the location $r=r_S$ is not the fixed point of the circular coordinate $\tau$. It forms a circle near the tip of the cigar, with the size (radius) at order $\varepsilon_q$.}
\end{figure}

In the previous section, we look for the metric of manifold $\hat{\mathcal{M}}_q$, which is then given by eq. (\ref{tildemetricq}), and identify $\tilde{\tau}$ with $\tilde{\tau}+2\pi$ in order to introduce a conical singularity with deficit angle $\Delta\phi=2\pi(1-1/q)$ at the fixed point $r=r_S$. It seems like we simply add the factor of $1/q^2$ and keep the periodicity $\tilde{\tau}\sim\tilde{\tau}+2\pi$ to impose the existence of a conical singularity at $r=r_S$. The question now is whether we could find a full metric which reduces to eq. (\ref{tildemetricq}) near the fixed point located at $r=r_S$. This step is important to make sure whether the manifold $\hat{\mathcal{M}}_q$ satisfies the equation of motion which is required since we have set $G_{\mu\nu}=0$ in eq.  (\ref{derivativeaction}).\\
\indent We may make a first guess by replacing the original black hole's inverse temperature by $\beta\rightarrow \beta q$, to set the periodicity of $\tilde{\tau}$ to be $2\pi q$ for the smooth geometry, and then set $\tilde{\tau}\sim\tilde{\tau}+2\pi$ to introduce the conical singularity. However, this gives rise to a Schwarzschild black hole with the radius of the horizon $r_S^{(q)}=2Mq$, which is larger than the original entangling surface located at $r_S=2M$ for $q>1$. Thus, we cannot expand the metric near $r=2M$ since it is not allowed in the Euclidean calculation (recall that we need to calculate region only at the radius larger than the black hole horizon, which now it becomes $r>r_S^{(q)}$). Furthermore,, simply imposing the periodicity of $\tilde{\tau}$ to be $2\pi q$ everywhere in the bulk while keeping the metric unchanged may give the correct result for the calculation of entanglement entropy \cite{Fursaev2006} but might leads to incorrect result when we consider the R\'enyi entropy for arbitrary $q$ \cite{Headrick2010} since the full spacetime might not solve the Einstein equation. Here, we want to make sure that the full geometry satisfies the equation of motion.\\
\indent Another way to solve the problem is by replacing $r_S\rightarrow r_S/q$ in the original (Euclidean) Schwarzschild metric, 
\begin{equation}
ds^2=\bigg(1-\frac{r_S}{rq}\bigg)d\tau^2+\frac{dr^2}{\big(1-\frac{r_S}{rq}\big)}+r^2d\Omega^2.\label{fullschwarzschild}
\end{equation}
This metric can be seen as a Schwarzschild metric with horizon located at $r=r_S/q$ and temperature $T=q/4\pi r_S$. By expanding the metric near $\xi=0$, where $\xi=r-r_S$ is again the distance to the fixed point, we have
\begin{equation}
f_q(r)=\bigg(1-\frac{1}{q}\bigg)+\frac{1}{r_S q}\xi\equiv\varepsilon_q+\frac{1}{r_S q}\xi,
\end{equation}
from keeping only the first term of $\xi$. Here, we get an extra term $\varepsilon_q$ which is of order $\mathcal{O}(1-1/q)$ and vanishes when $q=1$, i.e. $\varepsilon_1=0$. For $1<q<\infty$, we have $0<\varepsilon_q<1$. By keeping the value of $\xi$ fixed and expand the metric near $q=1$, i.e. by putting all the terms which include $\varepsilon_q$ as the subleading terms, we could have
\begin{equation}
ds^2=\frac{\xi}{r_Sq}d\tau^2+\frac{r_Sq}{\xi}d\xi^2+r_S^2 d\Omega^2+\mathcal{O}(\varepsilon_q)\;,
\end{equation}
which reduces to 
\begin{equation}
ds^2=\frac{\tilde{y}^2}{q^2}d\tilde{\tau}^2+d\tilde{y}^2+r_S^2 d\Omega^2+\mathcal{O}(\varepsilon_q),\label{tildeqaway}
\end{equation}
from the parameterizations $y^2=\xi$, $\tilde{y}^2=4r_Sq y^2$, and $\tilde{\tau}=\tau/2r_S$. Thus, the metric (\ref{fullschwarzschild}) seems to be the good candidate for $\hat{\mathcal{M}}_q$.\\
\indent The metric given by eq. (\ref{tildeqaway}) differs $\mathcal{O}(\varepsilon_q)$ away from the metric of $\hat{\mathcal{M}}_q$ near the horizon (\ref{tildemetricq}). Thus, to calculate the entanglement entropy, we first ignore the $\mathcal{O}(\varepsilon_q)$ from the fact that we will take the limit $q\rightarrow1$ in the end, then calculate the surface term in eq. (\ref{derivativeaction}). The difference of order $\mathcal{O}(\varepsilon_q)$ comes from the fact that the manifold $\hat{\mathcal{M}}_q$ requires the location $r=r_S$ to be the fixed point of the replica symmetry, instead of the new Schwarzschild horizon $r_S/q$. Indeed, we have the difference between the fixed point $r_S$ and the new Schwarzschild horizon $r_S/q$ to be of order $\mathcal{O}(\varepsilon_q)$. We cannot simply plug the metric (\ref{fullschwarzschild}) into $\partial_q I_E[\hat{\mathcal{M}}_q]$ in eq. (\ref{derivativeaction}) to calculate the entanglement entropy, since eq. (\ref{fullschwarzschild}) is not actually the metric for $\hat{\mathcal{M}}_q$. If we try to do so, the calculation diverges after taking the limit $q=1$ (see Appendix B). \\
\indent However, instead of directly calculating the full spacetime in eq. (\ref{fullschwarzschild}), we could first consider the metric perturbatively by keeping only the first subleading order of $\varepsilon_q$. The full Euclidean metric is then given by
\begin{equation}
ds^2=\bigg(1-\frac{r_S}{r}+\frac{r_S}{r}\varepsilon_q\bigg)d\tau^2+\bigg[\frac{1}{1-r_S/r}-\frac{r_S}{r(1-r_S/r)^2}\varepsilon_q\bigg]dr^2+r^2d\Omega^2.
\end{equation}
We can then write the full metric by $g_{\mu\nu}^{(q)}=\bar{g}_{\mu\nu}+\delta g_{\mu\nu}^{(q)}$, where $\bar{g}_{\mu\nu}$ is the background metric when $q=1$, and the perturbation vanishes at $q=1$, i.e. $\delta g_{\mu\nu}^{(1)}=0$. Using the inverse of the metric perturbation $\delta g^{\mu\nu(q)}=\bar{g}^{\mu\alpha}\bar{g}^{\nu\alpha}\delta g_{\mu\nu}^{(q)}$, we obtain $\delta g_{\mu\nu}^{(q)}$ and $\delta g^{\mu\nu(q)}$ be
\begin{align*}
\delta g_{\tau\tau}^{(q)}=\frac{r_S}{r}\varepsilon_q,\;\;\;\delta g_{rr}^{(q)}=-\frac{r_S}{r(1-r_S/r)^2}\varepsilon_q,\;\;\\\nonumber
\delta g^{\tau\tau(q)}=-\frac{r_S}{r(1-r_S/r)^2}\varepsilon_q,\;\;\delta g^{rr(q)}=\frac{r_S}{r}\varepsilon_q.
\end{align*}
We then use these information to calculate $\partial_q I_E[\hat{\mathcal{M}}_q]$ only to the leading term of $\varepsilon_q$,
\begin{equation}
\partial_qI_E[\hat{\mathcal{M}}_q]=\int_{r=r_S+\varepsilon}\frac{d^3x}{16\pi G_N}\sqrt{\gamma}\hat{n}_\rho\big(\bar{g}_{\mu\nu}\bar{\nabla}^\rho\partial_q\delta g^{\mu\nu(q)}-\bar{\nabla}_\mu\partial_q\delta g^{\mu\rho(q)}\big),\label{derivativeperturbedschwarzschild}
\end{equation}
where $\gamma$ is now the induced metric of a surface with $r=r_S+\varepsilon$, $\hat{n}_\rho$ is its unit normal vector, and all components with overbar denotes the calculations in the background metric. The Euclidean time coordinate $\tau$ is integrated from $0$ to $\beta$, where $\beta = 4\pi r_S$, to introduce the coordinate singularity at the origin.\\
\indent Direct calculation of eq. (\ref{derivativeperturbedschwarzschild}) gives
\begin{equation}
\partial_qI_E[\hat{\mathcal{M}}_q]=\frac{4\pi r_S^2}{4G_N q^2},
\end{equation}
which perfectly reproduces eq. (\ref{deractionresult}), and thus the entanglement entropy is again given by eq. (\ref{entropyarea}). This provide us an alternative way to calculate the entanglement entropy, without only relying on the near-horizon geometry. This also tell us that the metric for $\hat{\mathcal{M}}_q$ solves the equation of motion to the leading order of $\varepsilon_q=1-1/q$, or near $q=1$, as required by \cite{LewkowyczMaldacena2013}. Therefore, despite having no problem in calculating the entanglement entropy, a general extension to the calculation of the R\'enyi entropy for arbitrary $q$ still needs to be studied more. Illustrations for $\hat{\mathcal{M}}_q$ in both Euclidean and Lorentzian setting can be seen in figure \ref{fig:fullEuc} and figure \ref{fig:fullLor} respectively.
\section{Entanglement entropy of More General Static and Spherically Symmetric Black Holes}
\label{sec:3}
In this section, we consider entanglement in more general black hole solutions to the Einstein field equation. Generally, the Einstein field equation does not need to be vacuum, i.e. it can be represented by a non-zero energy-momentum tensor of various matter fields such as the electromagnetic fields, quintessential matter, and cosmological constant. A general static and spherically-symmetric black hole solution in 4 dimension can be written as
\begin{equation}
ds^2=-f(r)dt^2+\frac{dr^2}{f(r)}+r^2d\Omega^2,\label{staticsphericalbh}
\end{equation}
where $f(r)$ is a function that gives zero at the horizon, i.e. $f(r_H)=0$, and becomes infinite in the origin $f(0)\rightarrow\infty$. For asymptotically-flat black holes, we have another restriction which is $f(\infty)\rightarrow1$, but this is not the case for black holes in AdS or dS. In a Schwarzschild black hole, $f(r)$ is given by $f(r)=1-r_S/r$ and hence the horizon is given by $r_H=r_S$. The Euclidean metric of eq. (\ref{staticsphericalbh}) is
\begin{equation}
ds^2=f(r)d\tau^2+\frac{dr^2}{f(r)}+r^2d\Omega^2.\label{staticsphericaleuclidean}
\end{equation}
\indent We can obtain the near-horizon geometry of eq. (\ref{staticsphericaleuclidean}) by expanding the function $f(r)$ near $r=r_H$,
\begin{equation}
f(r)=f(r_H)+f'(r_H)\xi+\frac{1}{2}f''(r_H)\xi^2+...\;,
\end{equation}
where $f'(r_H)\equiv\frac{df}{dr}\big|_{r=r_H}$. The first term of the right hand side is zero by definition of the horizon $f(r_H)=0$ and $f'(r_H)\neq0$ for non-extreme black holes. By keeping only the first power of $\xi$, we obtain
\begin{equation}
ds^2=f'(r_H)\xi d\tau^2+\frac{d\xi^2}{f'(r_H)\xi}+r_H^2d\Omega^2+...\;.
\end{equation}
Using parameterizations similar to the ones in the previous section, i.e. $y^2=\xi$, $\tilde{y}=\frac{2}{f'(r_H)}y$, and $\tilde{\tau}=\frac{f'(r_H)}{2}\tau$, we obtain
\begin{equation}
ds^2=\tilde{y}^2d\tilde{\tau}^2+d\tilde{y}^2+r_H^2d\Omega^2+...\;,\label{nearhorizongeneral}
\end{equation}
which again an $\mathbb{R}^2\times \textbf{S}^2$ geometry near $\tilde{y}=0$.  To avoid coordinate singularity at $\tilde{y}=0$, we have $\tilde{\tau}\sim\tilde{\tau}+2\pi$ or $\tau\sim\tau+\beta$ where $\beta=4\pi/f'(r_H)$ is the temperature of the horizon.\\
\indent Notice a crucial point here regarding the function $f(r)$. We can only have an $\mathbb{R}^2\times \textbf{S}^2$ near-horizon geometry if the black hole under consideration is non-extreme. This is so because for an extreme black hole with $f'(r_H)=0$, we will instead have $\text{AdS}_2\times \textbf{S}^2$ geometry near the horizon. We focus on the $\mathbb{R}^2\times \textbf{S}^2$ near-horizon geometry so that it is straightforward to obtain $\hat{\mathcal{M}}_q$, i.e. by adding the parameter $q$ such that the metric becomes
\begin{equation}
ds^2=\frac{\tilde{y}^2}{q^2}d\tilde{\tau}^2+d\tilde{y}^2+r_H^2d\Omega^2+...\;,\label{replicamanifoldgeneral}
\end{equation}
while keeping $\tilde{\tau}\sim\tilde{\tau}+2\pi$. Again, by plugging it into eq. (\ref{derivativeaction}) this form of metric reproduces the area law of entanglement entropy (\ref{entropyarea}) with area given by $4\pi r_H^2$. \\
\indent In this case, we still use the surface term of (\ref{derivativeaction}) to calculate $\partial_qI_E[\hat{\mathcal{M}}_q]$ for more general static and spherically symmetric black holes even though the solution (\ref{staticsphericalbh}) may comes from Einstein-Hilbert action coupled with some matter action $I_E[g,\psi]$ which is given by
\begin{equation}
I_E[g,\psi]=\frac{1}{16\pi G_N}\int d^4x\sqrt{g}R+I_m[\psi],
\end{equation}
where $I_m[\psi]$ stands for a (Euclidean) matter action for some matter field $\psi$. This is so because in calculating the variation of the action with respect to the R\'enyi index $q$, we will obtain
\begin{equation}
\partial_qI_E[\hat{\mathcal{M}}_q]=\int_{\hat{\mathcal{M}}_q}\frac{d^4x}{16\pi G_N}\sqrt{g}[E_{\mu\nu}\partial_q g^{(q)\mu\nu}+E_\psi\partial_q\psi]+\text{surface term},
\end{equation}
which now $E_{\mu\nu}=0$ is the equation of motion which includes the energy-momentum tensir. The term including $E_\psi$ also vanishes due to the equation of motion for the matter field $\psi$ \cite{LewkowyczMaldacena2013, Dong_2016}. The surface term is identical to the one in eq. (\ref{derivativeaction}) and thus giving us the same result. We also impose a boundary condition such that the variation the action at infinity vanishes.
\\
\indent Notice that, similar to the Schwarzschild case, the near-horizon geometry is all we need in calculating the entanglement entropy. Therefore, we may only focus with the behavior of the function $f(r)$ near the horizon. Indeed, this mean that any asymptotic behavior of the function $f(r)$, i.e. the behavior of $f(r)$ for large $r$, does not contributes to the result stating that the entanglement entropy is proportional to the area of the horizon. However, the asymptotic behavior still important for considering the regions that we are about to calculate, especially for black holes in asymptotically AdS or dS spacetimes. This will be discussed more in the next section. In principle, any form of $f(r)$ with $f'(r_H)\neq0$ may work.\\
\indent In this way, the entanglement entropy obtained by calculating $\partial_qI_E[\hat{\mathcal{M}}_q]$ calculates the entanglement between degrees of freedom in two timelike regions of the maximally-extended black hole spacetimes separated by the horizon $r_H$. If we mention entangled quantum degrees of freedom, in this work, we always meant the hypothetical quantum gravitational degrees of freedom living in those black hole spacetimes by the semiclassical assumption explained in section 2.2. The area law of the entropy comes from strong short-range interactions near the horizon. Any long-range interaction would contributes as the sub-leading term or correction term to the entropy. However, in this work, we only discuss the calculation of the leading term which gives us the area-law entropy.\\
\indent The difference between this case and the Schwarzschild calculation lies in the number of horizon(s) that $f(r)$ can produce. The equation $f(r_H)=0$ can have more than one roots and thus more than one horizons.  This also follows from the fact that the black hole under consideration needs to be non-extreme, avoiding the case of a more general black hole with a single horizon. Furthermore, for black holes with multiple horizons, the value of $f(r)$ in some regions can be either positive or negative. Therefore, we have to be careful on which horizon should we use in calculating the entanglement entropy and which regions become our interest. We will tackle this problem by first considering the Reissner-Nordstr\"om black hole in the next subsection.\\\\
\textbf{General static and spherically symmetric case without near-horizon limit}\\
Before going into the calculation of the RN black holes, in this part, we calculate the entanglement entropy by considering the full metric $\hat{\mathcal{M}}_q$ of a more general static and spherically-symmetric black holes, following similar logic of Section \ref{sec:2.3}. The function $f(r)$ will be replaced by $f_q(r)$ with $f_1(r)=f(r)$, which now depends on the R\'enyi index $q$. Before knowing the explicit form for $f_q(r)$ for various black hole solutions, we first find what conditions need to be satisfied by $f_q(r)$. First, $f_q(r)$ is a black hole solution with $f_q(r_H^{(q)})=0$, where $r_H^{(q)}$ is the black hole horizon's radius which satisfies $r_H^{(1)}=r_H$. Second, we expect that near the entangling surface $r_H$, the function becomes
\begin{equation}
f_q(r)\approx f_q(r_H)+f'_q(r_H)\xi=\mathcal{O}(\varepsilon_q)+\frac{f'(r_H)}{q}\xi.\label{fqconstraint}
\end{equation}
The second equality follows from the fact that if we ignore the $\mathcal{O}(\varepsilon_q)$ term, we will have the near-entangling surface region as
\begin{equation}
ds^2=\frac{f'(r_H)}{q}\xi d\tau^2+\frac{q}{f'(r_H)}\frac{d\xi^2}{\xi}+r_H^2d\Omega^2+...\;,
\end{equation}
which using parameterizations $y^2=\xi$, $\tilde{y}^2=4qy^2/f'(r_H) $, and $\tilde{\tau}=f'(r_H)\tau/2$, we recover
\begin{equation}
ds^2=\frac{\tilde{y}^2}{q^2}d\tilde{\tau}^2+d\tilde{y}^2+r_H^2d\Omega^2+...\;.
\end{equation}
This is required because near $r=r_H$, the metric for $\hat{\mathcal{M}}_q$ needs to be in the form of eq. (\ref{replicamanifoldgeneral}) and keep $\tilde{\tau}\sim\tilde{\tau}+2\pi$ to introduce conical singularity with deficit angle $\Delta\phi=2\pi(1-1/q)$.\\
\indent We want to calculate $\partial_qI_E[\hat{\mathcal{M}}_q]$ by using the metric for a static and spherically-symmetric black hole with the replacement $f(r)\rightarrow f_q(r)$ perturbatively in $\varepsilon_q=1-1/q$. By expanding $f_q(r)$ near $\varepsilon_q=0$, we have
\begin{equation}
f_q(r)=f(r)+\partial_qf_q(r)\big|_{q=1}\varepsilon_q+...\;.
\end{equation}
Thus, the static and spherically symmetric black hole metric becomes
\begin{equation}
ds^2=\bigg[f(r)+\partial_qf_q(r)\big|_{q=1}\varepsilon_q\bigg]d\tau^2+\bigg[\frac{1}{f(r)}-\frac{\partial_q f_q(r)\big|_{q=1}}{f^2(r)}\varepsilon_q\bigg]dr^2+r^2d\Omega^2.\label{perturbedgeneral}
\end{equation}
The metric perturbation $\delta g_{\mu\nu}^{(q)}$ can be written as
\begin{align*}
\delta g_{\tau\tau}^{(q)}=\partial_q f_q(r)\big|_{q=1}\varepsilon_q,\;\;\delta g_{rr}^{(q)}=-\frac{\partial_q f_q(r)\big|_{q=1}}{f^2(r)}\varepsilon_q,\\
\delta g^{\tau\tau(q)}=\frac{\partial_q f_q(r)\big|_{q=1}}{f^2(r)}\varepsilon_q,\;\;\delta g^{rr(q)}=-\partial_q f_q(r)\big|_{q=1}\varepsilon_q.
\end{align*}
Using the perturbed metric in eq. (\ref{perturbedgeneral}), we obtain
\begin{align}
\partial_q I_E[\hat{\mathcal{M}}_q]&=\int_{r=r_H+\varepsilon}\frac{d^3 x}{16 \pi  G_N}\sqrt{\gamma}\hat{n}_\rho\big(\bar{g}_{\mu\nu}\bar{\nabla}^\rho\partial_q\delta g^{\mu\nu(q)}-\bar{\nabla}_\mu\partial_q\delta g^{\mu\rho(q)}\big)\\\nonumber
&=\frac{(4\pi r_H^2)}{4G_Nq^2}\times\frac{1}{f'(r_H)}\bigg[\frac{2}{r_H}\partial_q f_q(r_H)\big|_{q=1}+\partial_q f'_q(r_H)\big|_{q=1}\bigg],
\end{align}
Where we have used $\partial_r\partial_qf(r)=\partial_q f'(r)$, and the integral is performed on a hypersurface with constant radius $\varepsilon$ centered at $r_H$ and then take the limit $\varepsilon\rightarrow0$. On the other hand, from eq. (\ref{fqconstraint}), we have
\begin{equation}
\partial_q f'_q(r_H)=\partial_q\bigg[\frac{f'(r_H)}{q}\bigg]=-\frac{f'(r_H)}{q^2}.
\end{equation}
Hence, $\partial_q I_E[\hat{\mathcal{M}}_q]$ becomes
\begin{equation}
\partial_qI_E[\hat{\mathcal{M}}_q]=\frac{(4\pi r_H^2)}{4G_N q^2}\times\frac{1}{f'(r_H)}\bigg[\frac{2}{r_H}\partial_q f_q(r_H)\big|_{q=1}-f'(r_H)\bigg].\label{deractiongeneral}
\end{equation}
\indent The value of $\partial_qf_q(r_H)\big|_{q=1}$ needs to be determined from the explicit function of $f_q(r)$ for each black hole case. Nevertheless, we expect it to be $\partial_qf_q(r_H)\big|_{q=1}=r_Hf'(r_H)$, to reproduce the result in the form of eq. (\ref{deractionresult}), which is true for a Schwarzschild black hole.  In the following sections, we also see that it is true for a Reissner-Nordstr\"om and Schwarzschild-de Sitter black hole. The calculations of the entanglement entropy without near-horizon geometry for more general black holes also suggest that the manifold $\hat{\mathcal{M}}_q$ satisfies the equation of motion at the leading order of $\varepsilon_q$.\\
\indent The function $f_q(r)$ should depends on $q$ only at its parameter(s). For example, in Schwarzschild solution, the dependence on $q$ is present at the mass parameter, which is replaced by $M/q$ from the original mass $M$. This replacement does not change the black hole geometry, i.e., the Schwarzschild black hole geometry in $\mathcal{M}$ remains a Schwarzschild black hole geometry in the replica manifold $\hat{\mathcal{M}}_q$ with mass replaced by $M/q$. A form of 
\begin{equation}
f_q(r)=q-\frac{2M}{r},
\end{equation}
for example, cannot exist since the geometry is not Schwarzschild solution anymore. This is because we need to keep the black hole geometry and to make sure that the replica manifold $\hat{\mathcal{M}}_q$ satisfies the Einstein equation of motion. This requirement also applies to other black holes which is discussed in the next section.
\subsection{Charged Reissner-Nordstr\"om Black Hole}
\label{sec:3.1}
\begin{figure}
\begin{center}
\includegraphics[scale=1.3]{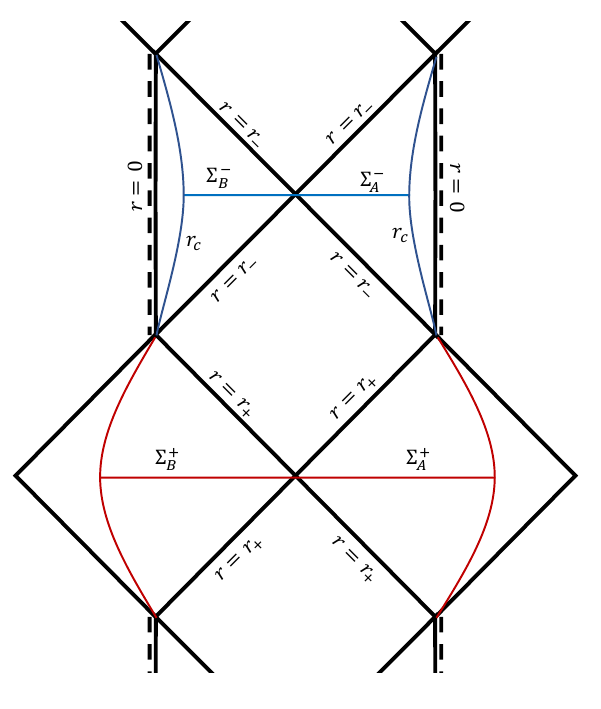}
\end{center}
\caption{Carter-Penrose diagram of a maximally-extended Reissner-Nordstr\"om spacetime. The blue curved line represents the cut-off sphere $r_c$ that isolates the true singularity while the red curved line represents arbitrary cut-off which can be taken to be infinite.}\label{fig:RNPenrose}
\end{figure}
\begin{figure}
\centering
\includegraphics[scale=0.7]{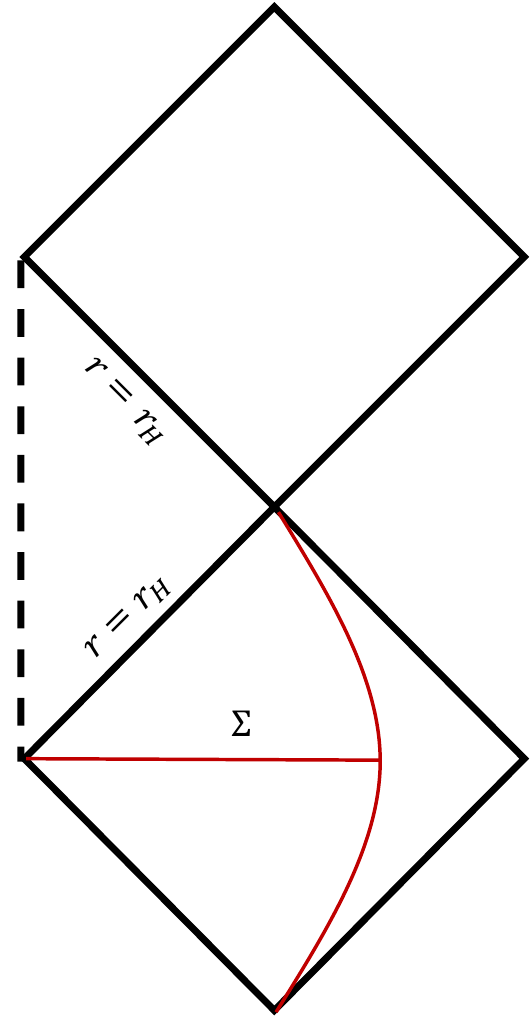}
\caption{Carter-Penrose diagram of an extreme Reissner-Nordstr\"om black hole. The slice $\Sigma$ is a "total" slice which represent a pure density matrix $\rho$ since now it is not separated into two subregions with the horizon in between.}\label{fig:extremeRN}
\end{figure}

A Reissner-Nordstr\"om black hole is a static and spherically symmetric solution of the Einstein-Maxwell equation which comes from the Einstein-Hilbert action coupled with Maxwell action given by
\begin{equation}
I_m=-\frac{1}{4}\int d^4x\sqrt{g}F^{\mu\nu}F_{\mu\nu},
\end{equation}
where $F_{\mu\nu}=\partial_\mu A_\nu-\partial_\nu A_\mu$ is the electromagnetic strength tensor and $A_\mu$ is the electromagnetic gauge field with
\begin{equation}
A_\mu dx^\mu=\frac{Q}{r}dt.
\end{equation}
From now on, we set $G_N=1$. The Reissner-Nordstr\"om solution is given by eq. (\ref{staticsphericalbh}) with
\begin{equation}
f(r)=1-\frac{2M}{r}+\frac{Q^2}{r^2},
\end{equation}
where $Q$ is the charge of the black hole. The horizon equation $f(r_H)=0$ yields to two roots given by
\begin{equation}
r_{\pm}=M\pm\sqrt{M^2-Q^2},
\end{equation}
which are called the inner $r_-$ and outer $r_+$ horizon ($r_-<r_+$). A non-extreme Reissner-Nordstr\"om black hole requires $M^2>Q^2$. The Carter-Penrose diagram for Reissner-Nordstr\"om black hole is depicted by figure \ref{fig:RNPenrose}.
\\
\indent To avoid coordinate singularity at the horizons $r_\pm$, we perform coordinate transformations to the Kruskal-Szekeres coordinates. However, in this case, we need two patches; the one which omits the coordinate singularity at $r_+$ while still having a coordinate singularity at $r_-$ and the other one which describes the opposite. Those two patches are given by (see, for example, \cite{griffiths_exact_2009})
\begin{equation}
ds^2=\frac{r_+r_-}{r^2}\bigg(\frac{r-r_-}{r_-}\bigg)^{1+\frac{r_-^2}{r_+^2}}\exp\bigg(-\frac{r_+-r_-}{r_+^2}r\bigg)(-dT_{+}^2+dR_+^2)+r^2d\Omega^2,
\end{equation}
and
\begin{equation}
ds^2=\frac{r_+r_-}{r^2}\bigg(\frac{r_+-r}{r_+}\bigg)^{1+\frac{r_+^2}{r_-^2}}\exp\bigg(\frac{r_+-r_-}{r_-^2}r\bigg)(-dT_-^2+dR_-^2)+r^2d\Omega^2.
\end{equation}
The coordinates $T_\pm,R_\pm$ satisfy
\begin{equation}
-T_\pm^2+R_\pm^2=(\pm1)\frac{2r_\pm^2}{(r_+-r_-)^2}\bigg(\frac{r-r_\pm}{r_\pm}\bigg)\bigg(\frac{r_\mp}{r-r_\mp}\bigg)^\frac{r_\mp^2}{2r_\pm}\exp\bigg(\frac{r_\mp-r_\pm}{r_\pm^2}r\bigg).
\end{equation}
However, in a Euclidean continuation, i.e. $T_\pm\rightarrow iT_{E\pm}$,  we have $r>r_+$ for $T_{E+},R_{E+}$ and $0<r<r_-$ for $T_{E-},R_{E-}$. In a Euclidean Reissner-Nordstr\"om spacetime, only regions with $f(r)>0$ is covered, i.e. $r>r_+$ and $0<r<r_-$. This also concludes that in any black hole solution with Euclidean metric given by eq. (\ref{staticsphericaleuclidean}), we only focus on regions with $f(r)>0$.\\
\indent We will have a standard cigar similar to the one for Schwarzschild case at $r>r_+$ and another cigar facing the opposite direction at $0<r<r_-$. We will denote these cigars as $\mathcal{M}^+$ and $\mathcal{M}^-$ respectively.  Both $\mathcal{M}^+$ and $\mathcal{M}^-$ can be described by the metric (\ref{staticsphericaleuclidean}) with $r>r_+$ and $0<r<r_-$ respectively. To avoid coordinate singularities at the tip both cigars (at $r=r_+$ for $\mathcal{M}^+$ and $r=r_-$ for $\mathcal{M}^-$), the periodicity of the Euclidean time need to be $\tau\sim\tau+\beta_+$ and $\tau\sim\tau+|\beta_-|$ for $r>r_+$ and $0<r<r_-$ respectively, where
\begin{equation}
\beta_{\pm}=\frac{4\pi}{f'(r_\pm)}.
\end{equation}
$\mathcal{M}^+$ and $\mathcal{M}^-$ can have different periodicity for the coordinate $\tau$ without any problem since those cigars are disconnected. While $\mathcal{M}^+$ has a similar shape compared to the Schwarzschild case since both have $f(\infty)\rightarrow1$, $\mathcal{M}^-$ have a different geometry. The radius of the cigar should diverge at $r=0$ since $|f(0)|\rightarrow\infty$. This is problematic since it is difficult to consider the periodic nature of the cigar at $r=0$. We cannot include the true singularity $r=0$  into our calculations. To avoid the problem, we could define a small but non-zero cut-off $r_c$ such that $f(r_c)$ might be large but finite.  In this case, $\mathcal{M}^-$ is defined for $r_c<r<r_-$, which means that we isolate the true singularity $r=0$ by a spherical cavity of radius $r_c$, and only consider the degrees of freedom outside the cavity up to the inner horizon $r_-$.\\
\begin{figure}
\centering
\includegraphics[scale=0.5]{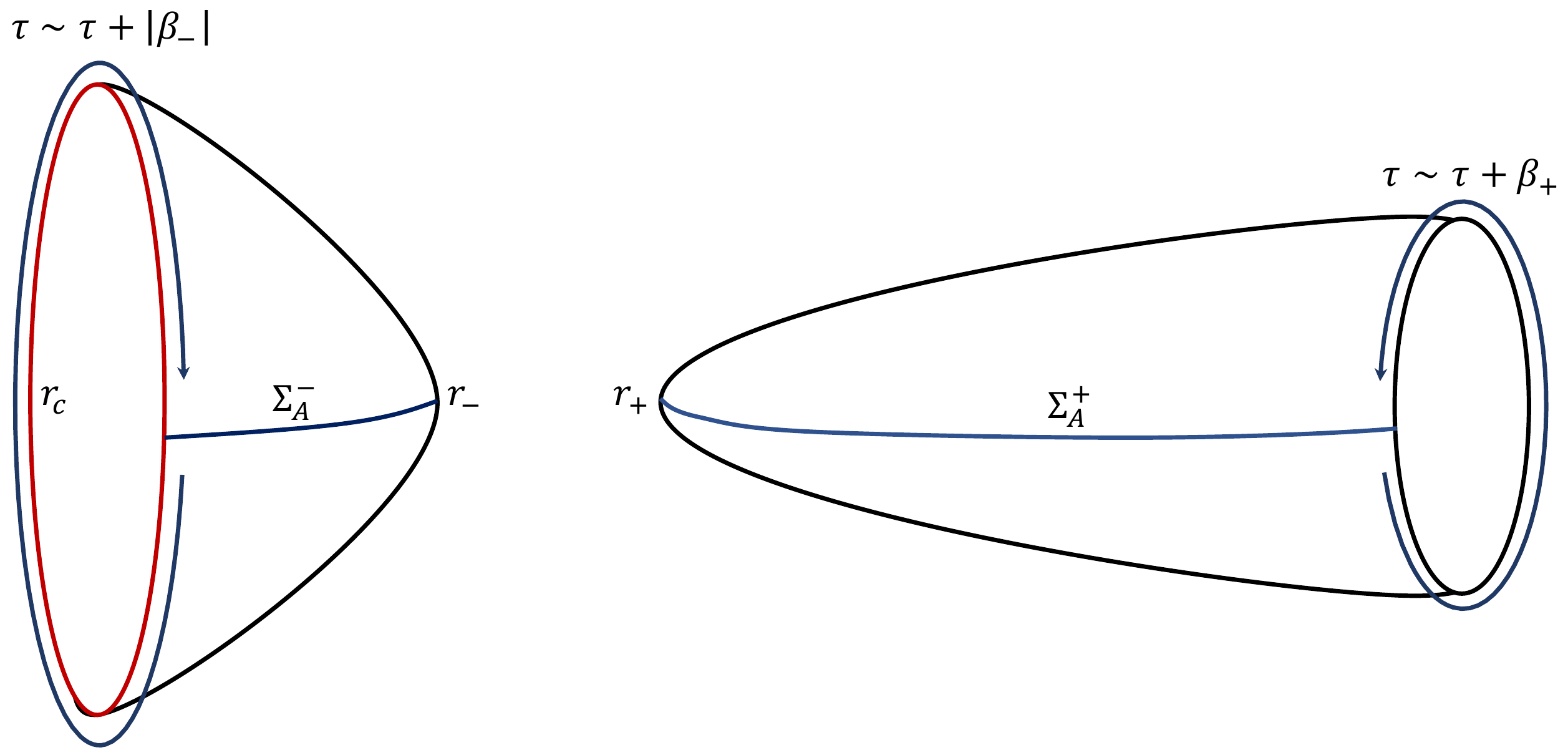}
\caption{Euclidean cigar geometry for the charged Reissner-Nordstr\"om black hole.}\label{fig:RNcigar}
\end{figure}
\indent We would like to calculate entanglement entropy in both $\mathcal{M}^\pm$. For $\mathcal{M}^+$, we pick a slice $\Sigma^+$ dividing $\mathcal{M}^+$ into two time-symmetric ($\tau=-\tau$) slices. The slice $\Sigma^+$ is then divided into two subregions $\Sigma^+=\Sigma^+_A\cup\Sigma^+_B$ such that
\begin{equation}
\Sigma_A^+=\{\tau,x^i|\tau=0,r\in[r_+,\infty)\},
\end{equation}
and
\begin{equation}
\Sigma_B^+=\{\tau,x^i|\tau=\beta_+/2,r\in[r_+,\infty)\}.
\end{equation}
To calculate the entanglement entropy of the subregion $\Sigma_A^+$, we calculate the von Neumann entropy of a reduced density matrix $\rho_A^+=\text{Tr}_B(\rho^+)$ via replica trick. This forces us to find the metric for the manifold $\hat{\mathcal{M}}_q^+$, which has a conical singularity at $r=r_+$ with deficit angle $\Delta\phi=2\pi(1-1/q)$.  The illustration of the manifold $\hat{\mathcal{M}}_q^+$ is given by figure \ref{fig:RNcigar} (right). The near-horizon metric is simply given by eq. (\ref{replicamanifoldgeneral}) with $\tilde{\tau}=\frac{f'(r_+)}{2}\tau$ and the radius of the $\textbf{S}^2$ part is $r_+$. Thus, plugging the metric to eq. (\ref{derivativeaction}) gives us the area law
\begin{equation}
S_A^+=\pi r_+^2.
\end{equation}
\indent Notice that the formula of $\partial_q I_E[\hat{\mathcal{M}}_q]$ is still given by eq. (\ref{derivativeaction}) since we also have the same boundary condition at infinity, similar to the Schwarzschild calculation. Physical interpretation of the leading term of the entanglement entropy is also similar to the Schwarzschild counterpart. However, when proceeding our calculations to the entanglement entropy at the inner horizon $r_-$, there will be several distinction from the one in $r_+$. We need to impose some boundary conditions at $r_c$, i.e. near the true singularity, to vanish. The metric near the sphere with radius $r=r_c$ behaves as
\begin{equation}
ds^2\approx\frac{Q^2}{r^2}d\tau^2+\frac{r^2}{Q^2}dr^2+ r^2d\Omega^2.\label{bcRN}
\end{equation}
Thus, if the boundary term at this surface vanishes, we require this metric to be independent of the parameter $q$, which implies that the charge parameter should be constant with respect to $q$. \\
\indent We need such boundary condition at $r_c$ since the physical interpretation of the entanglement entropy requires so. The entangled degrees of freedom away from the entangling surface located at $r_-$, i.e. near the boundary $r_c$, should not contributes to the leading term of the entanglement entropy, at least when $r_c\ll r_-$. One may ask what if we have a small inner horizon such that $r_c$ cannot be approximated by $r_c\ll r_-$. In this case, since $r_-=M-\sqrt{M^2-Q^2}$, we only have small $r_-$ in the limit of $Q\rightarrow 0$. However, in this limit, the metric reduces to the ordinary Schwarzschild spacetime, which does not have inner horizon at all. Hence, there is no entangled degrees of freedom to be considered anyway.\\
\indent Next, we consider the entanglement entropy in $\mathcal{M}^-$. Since $f'(r_-)<0$, we write the line element of the Euclidean near-horizon geometry of $\mathcal{M}^-$ to be
\begin{equation}
ds^2=|f'(r_-)|(-\xi)d\tau^2+\frac{d\xi^2}{|f'(r_-)|(-\xi)}+r_-^2d\Omega^2+...\;.
\end{equation}
However, $-\xi=r-r_-$ is now positive since we are working in the region $r_c<r<r_-$ and thus the parameterization $\tilde{y}=\frac{2}{|f'(r_-)|}\sqrt{-\xi}$ is real and therefore it is harmless. Hence, we recover the line element in eq. (\ref{nearhorizongeneral}) with $\tilde{\tau}=\frac{|f'(r_-)|}{2}\tau$ and $\tau\sim\tau+|\beta_-|$ as mentioned earlier. We also define a slice $\Sigma^-=\Sigma^-_A\cup\Sigma^-_B$ such that 
\begin{equation}
\Sigma_A^-=\{\tau,x^i|\tau=0,r\in[r_c,r_-]\},
\end{equation}
and
\begin{equation}
\Sigma_B^-=\{\tau,x^i|\tau=|\beta_-|/2,r\in[r_c,r_-]\}.
\end{equation}
Similar to the previous case, calculating the entanglement entropy of a reduced density matrix $\rho_A^-$ means plugging in the line element of $\hat{\mathcal{M}}_q^-$, which is nothing but eq. (\ref{replicamanifoldgeneral}) with $\tilde{\tau}=\frac{|f'(r_-)|}{2}\tau$ to eq. (\ref{derivativeaction}).  Since now the radius of the $\textbf{S}^2$ part is $r_-$, we recover the area law for the inner horizon,
\begin{equation}
S_A^-=\pi r_-^2.
\end{equation}
\indent We would like to make several remarks to the result of entanglement entropy in the maximally-extended Reissner-Nordstr\"om spacetime. $S_A^+$ calculates the entanglement entropy of the quantum gravitational degrees of freedom in $\Sigma_A^+$ while $S_A^-$ computes the ones in $\Sigma_A^-$. Each of them captures the correlations which are strong near each horizon and therefore produce the area law as the leading term of the entanglement entropy. Since $\Sigma^+$ and $\Sigma^-$ (and also $\mathcal{M}^+$ and $\mathcal{M}^-$) are completely disconnected in the Euclidean setting, we cannot see them as one and consider the total entropy as the sum of both entropy, i.e. $S_A^{\text{total}}\neq S_A^++S_A^-$ even though both inner $r_-$ and outer $r_+$ radius might be correlated in some ways since both are functions of $M$ and $Q$ simultaneously.  Such correlation might affect the thermodynamics of Reissner-Nordstr\"om black hole (see, for example, \cite{Volovik2021}) but this does not seem so when considering the entanglement entropy.\\
\indent We can also see this through the conformal diagram of the Reissner-Nordstr\"om black hole.  As can be seen in figure \ref{fig:RNPenrose}, it is clear that $\Sigma^+$ and $\Sigma^-$ does not lie in a same slice. When calculating the entanglement entropy of a subregion, we always consider subregions which are in a same (Cauchy) slice of constant time. Therefore, we see this problem as counting entanglement in two different (and separated) regions $\mathcal{M}^+$ and $\mathcal{M}^-$, giving two different entanglement entropy $S_A^+$ and $S_A^-$. This pictorial description can also help us in visualizing entanglement entropy in an extreme Reissner-Nordstr\"om black hole. In such limit, i.e. $r_-=r_+$ (or $M^2=Q^2$), the conformal diagram is depicted by figure (\ref{fig:extremeRN}). 
In that diagram, we cannot draw a single slice $\Sigma$-like surface which represent a total slice of two subregions separated by a horizon. Therefore, aside from the difficulty of calculating the entanglement entropy when $f'(r_H)=0$, entanglement entropy for extreme black hole is rather confusing. Reissner-Nordstr\"om black holes with naked singularity ($M^2<Q^2$) also leads to the similar conclusion.\\\\
\textbf{Remarks on the Extreme Reissner-Nordstr\"om Black Hole}\\
\indent If we put aside the difficulty of choosing a proper slice for calculating the entanglement entropy in an extreme black hole, mathematically, we are still able to calculate the entanglement entropy by first considering a near-extreme condition to the near-horizon geometry and take the extreme limit later on. Suppose that we are calculating the entanglement entropy of the outer horizon $r_+$. In a near-extreme condition, the outer horizon can be written as
\begin{equation}
r_+=r_H+\delta r_H,\label{nearextremeRNradius}
\end{equation}
where $r_H=M$ is the extreme black hole radius. If we are expressing eq. (\ref{nearextremeRNradius}) with some parameter $\zeta^2=M-Q$ and taking only the first order of $\zeta$, we obtain $\delta r_H=\sqrt{2M}\zeta$. In the limit $\zeta\rightarrow0$, we recover the extreme limit $Q\rightarrow M$. Thus, we can expand $f'(r_+)$ up to the first order of $\zeta$ and obtain
\begin{equation}
f'(r_+)\approx\frac{df'(r_+)}{dr_+}\bigg|_{r_+=r_H}\delta r_H=\frac{df'(Q)/dQ}{dr_+(Q)/dQ}\bigg|_{Q=M}\delta r_H=\sqrt{\frac{2}{M^3}}\zeta.
\end{equation}
Note that the value of $M$ is fixed and $Q$ is treated as the continuous variable,  and then we take the extreme limit $Q\rightarrow M$. In this case, the temperature of the black hole is given by
\begin{equation}
T_{\text{near-ext.}}\approx\frac{\zeta}{\pi\sqrt{8M^3}},
\end{equation}
where in the extreme limit we recover $T_{\text{ext.}}=0$.\\
\indent In this case, we are able to construct the near-horizon metric of manifold $\hat{\mathcal{M}}_q$ for a near-extreme Reissner-Nordstr\"om black hole, which is given by
\begin{equation}
ds^2=\frac{\tilde{y}^2}{q^2}d\bigg(\frac{\zeta}{\sqrt{2M^3}}\tau\bigg)^2+d\tilde{y}^2+r_+^2d\Omega^2+...\;.\label{replicametricRNnearext}
\end{equation}
Here we have $\tilde{y}=\frac{\sqrt{2M^3}}{\zeta}\sqrt{r-r_+}$ and we keep $\tilde{\tau}=\frac{\zeta}{\sqrt{2M^3}}\tau$ to be periodic with periodicity $2\pi$ (in order to avoid conical singularity when $q=1$). Hence, the Euclidean time is periodic with $\tau\sim\tau+\frac{\pi\sqrt{8M^3}}{\zeta}$. Up to this point, nothing is problematic since we keep $\zeta$ to be small but non-zero since in such case, the near-horizon geometry (when $q=1$) is still $\mathbb{R}^2\times\textbf{S}^2$. In eq. (\ref{replicametricRNnearext}), we also keep the radius of the $\textbf{S}^2$ part to be $r_+$ but at the moment, it is given by eq. (\ref{nearextremeRNradius}). Plugging this metric into eq. (\ref{derivativeaction}) and integrating $\tilde{\tau}$ from $0$ to $2\pi$ as usual gives us the entanglement entropy
\begin{equation}
S^{\text{near-ext.}}_A=\pi r_+^2,\label{entropyextremeplus}
\end{equation}
which becomes
\begin{equation}
S_A^{\text{ext.}}=\pi M^2,\label{extremeRNentropy}
\end{equation}
in the extreme limit. Thus, the entanglement entropy in an extreme Reissner-Nordstr\"om black hole is obtained without any issue and again we recover the area law of the entanglement entropy which is given by the area of the extreme horizon $r_H=M$. Calculations involving the inner horizon $r_-$, i.e. with with $r_-=r_H-\delta r_H$, gives us the same result. Both calculations from the outer horizon $r_+$ and inner horizon $r_-$ continuously reduces to eq. (\ref{extremeRNentropy}) in the extreme limit.\\
\indent An important question to ask is what the entropy of an extreme black hole in eq. (\ref{extremeRNentropy}) calculates. In the near-extreme limit, we still have a replica manifold $\hat{\mathcal{M}}_q$ with periodic Euclidean time coordinate where the periodicity is finite and is given by $\frac{\pi\sqrt{8M^3}}{\zeta}$. However, if we continue to the extreme limit with $\zeta\rightarrow0$, the Euclidean time coordinate becomes non-periodic. In this situation, we cannot have a replica manifold $\hat{\mathcal{M}}_q$ since the manifold $\hat{\mathcal{M}}_q$ requires the existence of a conical singularity at the origin with deficit angle $\Delta\phi=2\pi(1-1/q)$. Such conical singularity cannot exist if the angular coordinate becomes non-periodic and hence we cannot calculate the entanglement entropy from eq. (\ref{derivativeaction}).\\
\indent For an extreme RN black hole, the proper distance to the horizon is infinite, which can be seen from the line element of an extreme RN black hole with $Q=M$,
\begin{equation}
ds^2=-\bigg(1-\frac{M}{r}\bigg)^2dt^2+\frac{dr^2}{\big(1-\frac{M}{r}\big)^2}+r^2d\Omega^2.
\end{equation}
Therefore, when we construct a constant-time slice $\Sigma$ outside the horizon in the Euclidean setting, the horizon point $r=r_H$ is not included. Furthermore, in the Euclidean spacetime with $t=i\tau$, the Euclidean time coordinate $\tau$ does not have a specific value of periodicity to remove a conical singularity at the origin, since the origin itself is not included. We may fix the Euclidean time coordinate $\tau$ to be periodic with periodicity $\beta$, for any value of $\beta$, and the geometry becomes $\textbf{S}^1\times\mathbb{R}\times\textbf{S}^2$. Next, we define a tortoise-like coordinate
\begin{equation}
r_*=M\ln|r-M|+r,
\end{equation}
where the slice $\Sigma$ can now be written as
\begin{equation}
\Sigma=\{\tau,x^i|\tau=0,r_*\in(-\infty,\infty)\}.
\end{equation}
If we consider degrees of freedom outside the horizon, which is on the slice $\Sigma$, and construct a density matrix associated with $\Sigma$, the density matrix is already pure since it does not come from a partial trace, or a reduction, of a "total" density matrix. Therefore, the von Neumann entropy associated with $\rho$ becomes zero. Notice that this is true since we cannot construct a full slice $\Sigma$ where the degrees of freedom beyond (inside) the degenerate horizon is included. If such slice exist, it will be disconnected since the points located at the horizon is not included. This can also be seen from the conformal diagram of an extreme RN black hole, as explained earlier. Thus, we conclude that there is no notion of entanglement in an extreme RN spacetime, since the total slice outside the horizon represents a pure density matrix, and hence the entanglement entropy vanishes.\\
\indent The fact that the entanglement entropy of an extreme black hole is zero agrees with the thermodynamic entropy of a black hole from semiclassical calculations, which have been discussed earlier in \cite{Hawking1995,Carroll2009,Howard2013}. The discontinuity of the entropy of a non-extreme and extreme black hole comes from the fact that both of them can be considered as two totally different objects since we cannot continuously lower the mass of a non-extreme RN black hole into an extreme one \cite{Hawking1995}. There is also a discrepancy between the entropy calculation of an extreme black hole from semiclassical calculation and from counting the microstate of a CFT dual of the near-horizon $AdS_2\times \textbf{S}^2$ geometry since the latter gives an entropy which is finite and proportional to the black hole area \cite{Strominger1998}. However, the calculations done in \cite{Strominger1998} actually computes the entropy of a $AdS_2\times \textbf{S}^2$ spacetime,which is non-zero, and not the entropy of the actual extreme black hole, which is zero, even though the extreme black hole spacetime can be locally (near the horizon) described by $AdS_2\times \textbf{S}^2$ geometry \cite{Carroll2009}. Furthermore, perhaps we may see a non-extreme and extreme RN black hole as two spacetime having two different "phase", undergoing a phase transition at the critical temperature $T_c=0$ \cite{Pavan1991}.  This can be seen since the rate of change of the entropy if we vary the charge $Q$ diverges when the extreme limit $Q\rightarrow M$ is reached.  However, this will not be discussed in detail here and will be presented elsewhere.\\\\
\textbf{Calculations Without Near-Horizon Geometry}\\
In this part of this section, we calculate the entanglement entropy of a Reissner-Nordstr\"om black hole without near-horizon geometry, i.e. by finding the explicit value of $\partial_q f_q(r_H)\big|_{q=1}$ in eq. (\ref{deractiongeneral}). At first, consider the case for the entanglement entropy of the outer horizon $r_+$. The function $f(r)$ for a Reissner-Nordstr\"om black hole is controlled by two parameters which are its mass $M$ and charge $Q$. However, due to the boundary condition in eq. (\ref{bcRN}), $Q$ cannot be a function of the R\'enyi index $q$. Therefore, the only parameter changed is the mass $M$, which is replaced by $M\rightarrow M_q$ with $M_1=M$, and the function $f_q(r)$ can be written as
\begin{equation}
f_q(r)=1-\frac{2M_q}{r}+\frac{Q^2}{r^2}.\label{frqRN}
\end{equation}
\indent Following the condition
\begin{equation}
f'_q(r_+)=\frac{2M_q}{r_+^2}-\frac{2Q^2}{r_+^3}=\frac{f'(r_+)}{q},
\end{equation}
we have
\begin{equation}
M_q=\frac{r_+^2}{2}\bigg(\frac{f'(r_+)}{q}+\frac{2Q^2}{r_+^3}\bigg).\label{massqRN}
\end{equation}
The derivative of $M_q$ with respect to $q$ gives
\begin{equation}
\frac{dM_q}{dq}=-\frac{r_+^2f'(r_+)}{2q^2},\label{derMqRN+}
\end{equation}
and therefore the derivative of $f_q(r)$ with respect to $q$ gives
\begin{equation}
\partial_qf_q(r)=\frac{\partial_qf_q(r)}{\partial M_q}\frac{dM_q}{dq}=\frac{r_+^2f'(r_+)}{rq^2}.
\end{equation}
This is precisely what we expect since $\partial_qf_q(r_+)\big|_{q=1}=r_+f'(r_+)$, and the calculation of $\partial_qI_E[\hat{\mathcal{M}}_q]$ reproduces eq. (\ref{deractionresult}) with radius replaced by $r_+$. However, if we set the charge to be the function of $q$ instead of $M$, the result is incorrect. Thus, the identification in eq. (\ref{frqRN}) seems to be correct.\\
\indent Next, we investigate whether the new horizon $r_+^{(q)}$ lies inside or outside the original horizon $r_+$. Since $f'(r_+)$ is positive, from eq. (\ref{massqRN}), if the R\'enyi index is increased, then $M_q$ is decreased and since $q>1$, this means that $M_q<M$. From $f_q(r_+^{(q)})=0$ and $r_+^{(q)}$ being the positive root, i.e.
\begin{equation}
r_+^{(q)}=M_q+\sqrt{M_q^2-Q^2},
\end{equation}
decreasing $M_q$ opt to decreasing $r_+^{(q)}$ as well. However, the mass $M_q$ cannot be decreased too much because if it approaches $Q$, the black hole solution approach the extreme condition. Fortunately, in the limit $q\rightarrow\infty$, we have
\begin{equation}
M_\infty=\frac{Q^2}{r_+}=\frac{Q^2}{M+\sqrt{M^2-Q^2}},
\end{equation}
which is always larger than $Q$ and is equal to $Q$ only in the extreme limit of the original black hole, $M=Q$. Therefore, the value of $r_+^{(q)}$ is always smaller than $r_+$ for any $q$.\\
\indent For the inner horizon $r_-$, since $f'(r_-)<0$, the mass $M_q$ in eq. (\ref{massqRN}) becomes
\begin{equation}
M_q=\frac{r_-^2}{2}\bigg(-\frac{|f'(r_-)|}{q}+\frac{2Q^2}{r_-^3}\bigg),\label{MqRN-}
\end{equation}
and its derivative with respect to $q$ gives
\begin{equation}
\frac{dM_q}{dq}=\frac{r_-^2|f'(r_-)|}{q^2}.\label{derMqRN-}
\end{equation}
Thus, in this $r_-$ case, the derivative $\partial_q f_q(r)$ gives
\begin{equation}
\partial_q f_q(r)=\frac{\partial f_q(r)}{\partial M_q}\frac{dM_q}{dq}=-\frac{r_-^2|f'(r_-)|}{rq^2},
\end{equation}
and hence $\partial_qf_q(r_-)\big|_{q=1}=-r_-| f'(r_-)|$, which gives the correct result of $\partial_qI_E[\hat{\mathcal{M}}_q]$ as in eq. (\ref{deractionresult}) with radius replaced by $r_-$.
Notice that the derivative of the mass in eq. (\ref{derMqRN-} is identical to eq. (\ref{derMqRN+}) since $f'(r_\pm)$ can be written as $f'(r_\pm)=\frac{r_\pm-r_\mp}{r_\pm^2}$. However, the mass itself, i.e. between eq. (\ref{massqRN}) and eq. (\ref{MqRN-}), seems different while they should be the same. The mass $M_q$ can be written as
\begin{equation}
M_q=\frac{M}{q}+\frac{Q^2}{r_+}\varepsilon_q,
\end{equation}
or
\begin{equation}
M_q=\frac{M}{q}+\frac{Q^2}{r_-}\varepsilon_q.
\end{equation}
One can easily see that the mass only differs by $\varepsilon_q$, which in our case, we keep it to be small. The additional term involving $\varepsilon_q$ in the mass keeps the extreme limit (and even naked singularity) away if we increase $q$ up to $q\rightarrow\infty$, even though we only work in the regime where $q$ is near $1$. The value of $M_q$ also concludes that $r_-^{(q)}$, which is given by
\begin{equation}
r_-^{(q)}=M_q-\sqrt{M_q^2-Q^2},
\end{equation}
is always larger than $r_-$ for any $q$.
\subsection{Black Holes with Cosmological Horizon}
\label{sec:3.2}
\begin{figure}
\centering
\includegraphics[scale=0.6]{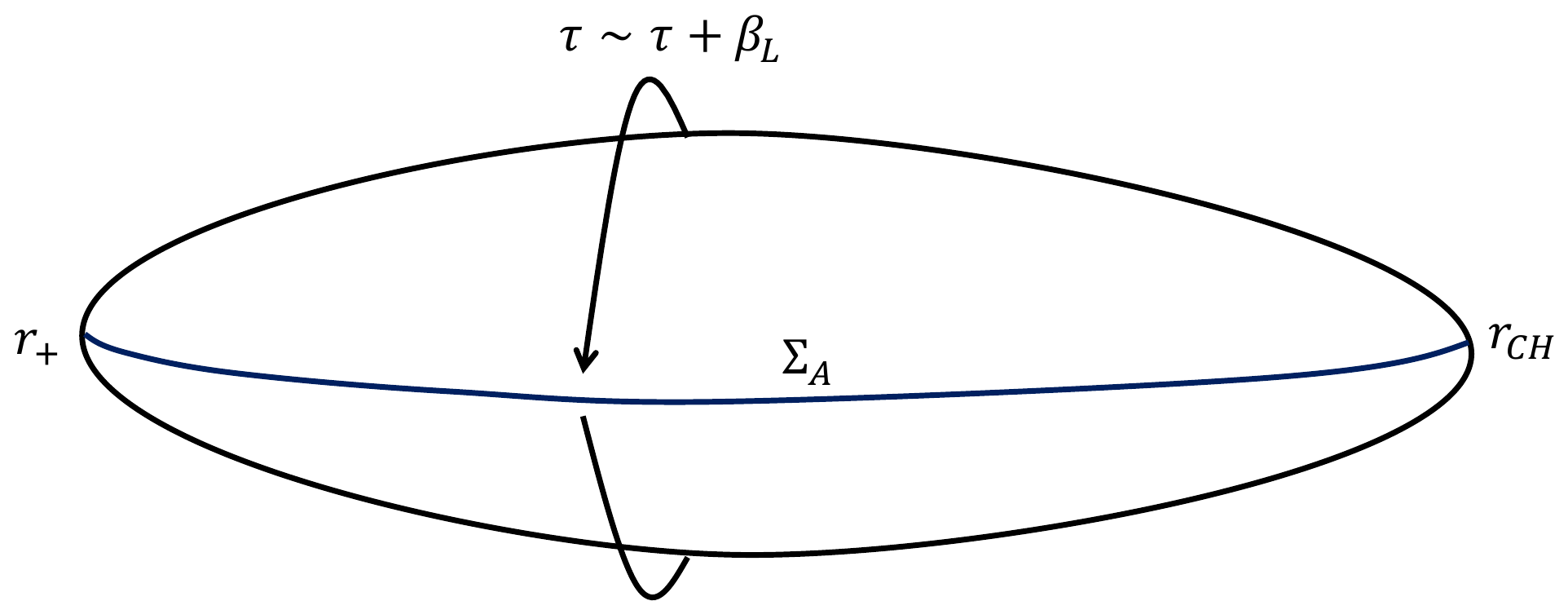}
\caption{The cigar geometry which represent a Euclidean Schwarzschild-de Sitter spacetime (or the region $r_+<r<r_{CH}$ of a RNdS spacetime). In the Lukewarm case, there is no conical singularity at the tips of the cigar, which are located in $r_+$ and $r_{CH}$.}\label{fig:SdSLukewarmCigar}
\end{figure}

\begin{figure}
\centering
\includegraphics[scale=0.55]{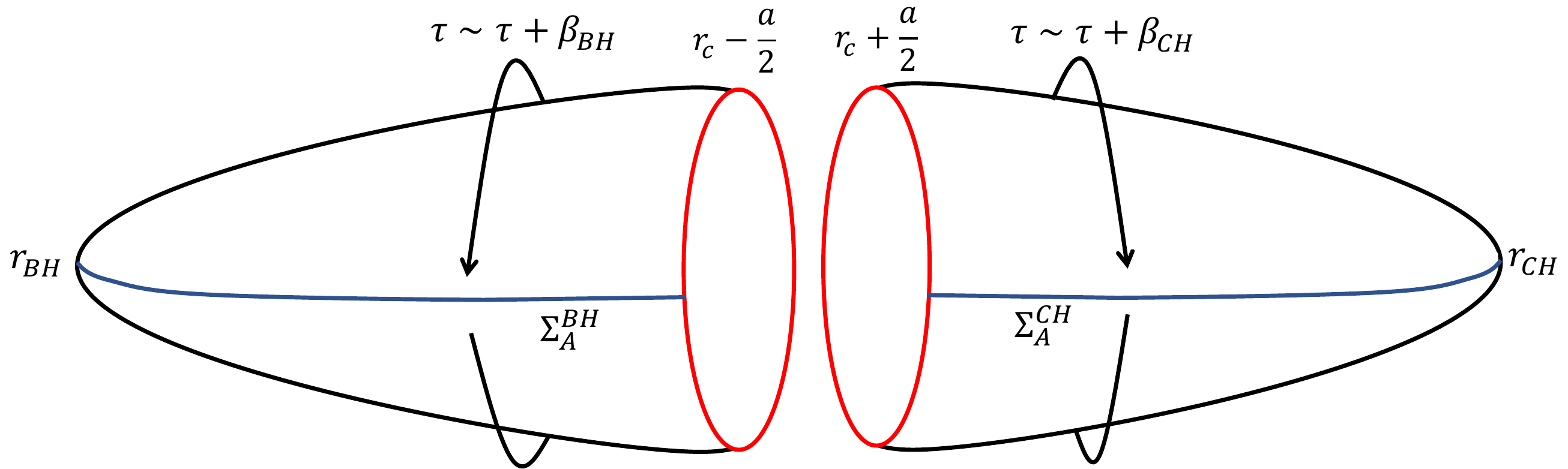}
\caption{Illustration for both $\hat{\mathcal{M}}_q^{BH}$ (left) and $\hat{\mathcal{M}}_q^{CH}$ (right). In this case, either left or right geometry have their on periodicity and hence there is no conical singularity in the original $q=1$ manifold.}\label{fig:SdSseparatecigar}
\end{figure}

In previous section, we study how to calculate entanglement entropy for black holes with inner horizon using the Reissner-Nordst\"om solution as the example. In this section, we study how to calculate entanglement entropy in black holes with cosmological horizon. In this case, the black hole solution can also have two (or more) horizons. We want to investigate the difference that arise in the calculation involving black holes with cosmological horizon and the ones without it. Some difficulties might arise if the black hole horizon and the cosmological horizon is not in a thermal equilibrium. As the first example, we consider the static and spherically symmetric solution to the Einstein equation with a positive cosmological constant $\Lambda$ or the Schwrarzschild-de Sitter spacetime.\\
\begin{figure}
\centering
\includegraphics[scale=0.7]{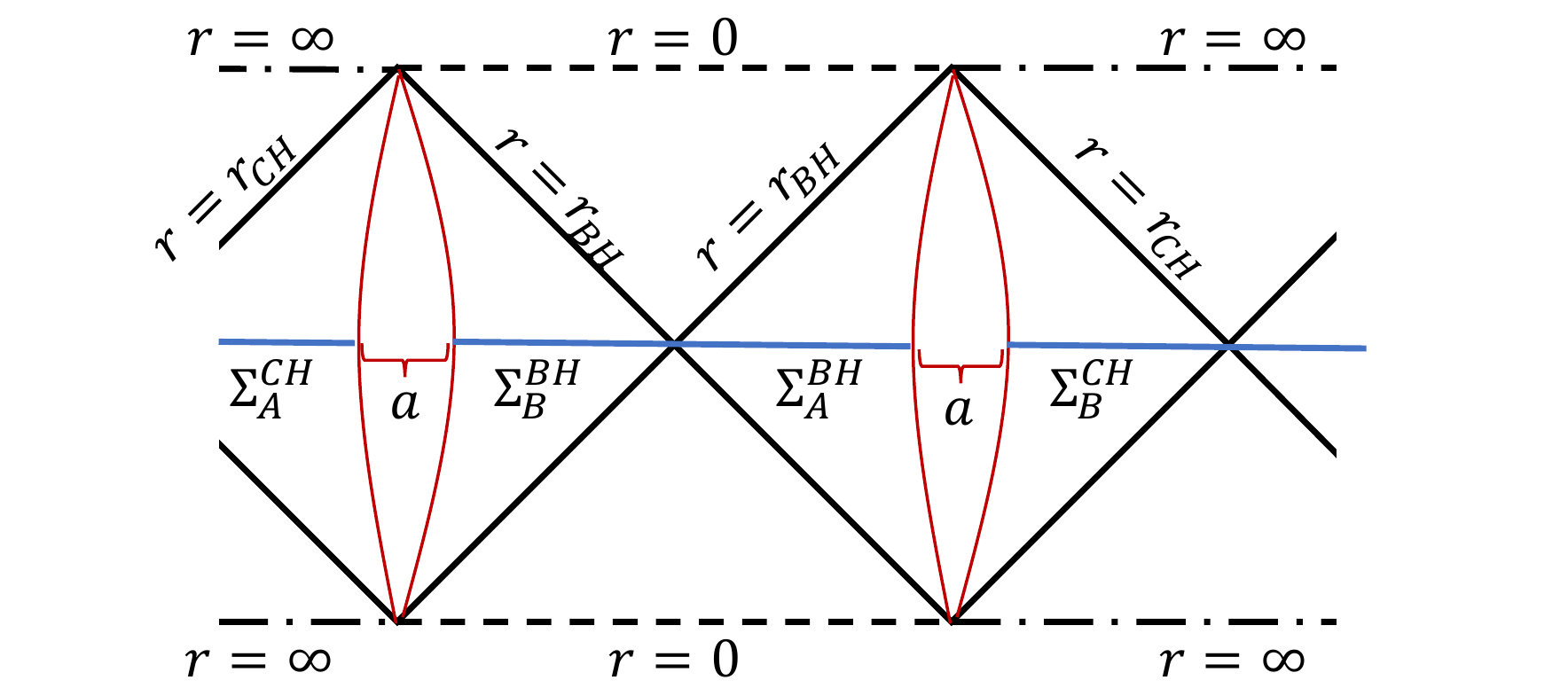}
\caption{Carter-Penrose diagram of a Schwarzschild-de Sitter spacetime. The rightmost part of the geometry can be seen as the identification to the leftmost part.  A "thin shell" with thickness $a$ separates the regions which is close to the black hole horizon $r_{BH}$ and the cosmological horizon $r_{CH}$.}\label{fig:SdSPenrose}
\end{figure}
\indent The Schwrazschild-de Sitter (SdS) spacetime is given by the metric in the form of eq. (\ref{staticsphericalbh}) with (again we use $G_N=1$)
\begin{equation}
f(r)=1-\frac{2M}{r}-\frac{r^2}{L^2},
\end{equation}
where the $L$ is the de Sitter radius which is related to the cosmological constant by $\Lambda=3/L^2$. The causal structure of this spacetime can be seen in figure \ref{fig:SdSPenrose}. Since $f(r)=0$ is now a cubic equation, we should have three solutions. However, only two solutions that are real and positive, named the black hole horizon $r_{BH}$ and the cosmological horizon $r_{CH}$. The black hole horizon is always the smaller solution and hence we have $r_{BH}<r_{CH}$. The value of $f(r)$ is only positive in the region $r_{BH}<r<r_{CH}$ and therefore we only work in such region when considering the Euclidean solution. Initially, we have $f(r_{BH})=0$ and it gradually increases when we increase $r$, until it decreases again at some point and finally going back to zero at $r_{CH}$. Thus the cigar geometry of SdS spacetime is given by a cucumber-like shape as can be seen in figure \ref{fig:SdSLukewarmCigar}.\\
\indent The cigar geometry of the SdS black hole will have at least one conical singularity at either $r=r_{BH}$ or $r=r_{CH}$ and cannot be removed simultaneously if the black hole horizon and the cosmological horizon is not in a thermal equilibrium. This is so because the near-horizon geometry near the black hole horizon is given by
\begin{equation}
ds^2=\tilde{y}^2d\bigg(\frac{f'(r_{BH})}{2}\tau\bigg)^2+d\tilde{y}^2+r_{BH}^2d\Omega^2 + ...\;,
\end{equation}
while the near-horizon geometry near the cosmological horizon is given by
\begin{equation}
ds^2=\tilde{y}^2d\bigg(\frac{|f'(r_{CH})|}{2}\tau\bigg)^2+d\tilde{y}^2+r_{CH}^2 d\Omega^2+ ...\;.
\end{equation}
If we want to remove the conical singularity at $r=r_{BH}$ by imposing the periodicity of $\tau$ to be $\tau\sim\tau+\frac{4\pi}{f'(r_{BH})}$, since this should apply globally throughout the cigar, conical singularity will appear at $r=r_{CH}$ because $f'(r_{BH})\neq |f'(r_{CH})|$. On the other hand, we will have conical singularity at $r=r_{BH}$ if we remove the one which is located at $r=r_{CH}$. Since for this kind of black hole $f'(r_H)$ is proportional to the surface gravity of the horizon, and hence proportional to the temperature of the horizon, this happen if the two horizons do not share the same temperature. In a SdS black hole, $f'(r_{BH})=|f'(r_{CH})|$ is satisfied if $r_{BH}=r_{CH}$ which is nothing but the Nariai limit. Since this is an extreme limit for SdS black holes, the case is not be considered in this work.\\
\indent Due to this, we cannot have a single Euclidean manifold representing a non-extreme SdS black hole cigar which is free from conical singularities. However, if we "cut" the cigar into two disconnected halves and give each of them their own periodicity of $\tau$, then we will not have any conical singularity. To do so, we add a cut-off $r_c$ in the region between $r_{BH}$ and $r_{CH}$ which satisfy $r_{BH}\ll r_c\ll r_{CH}$ and consider the region $r_{BH}<r<r_c$ and $r_c<r<r_{BH}$ separately. We will call manifold of the former region as $\mathcal{M}^{BH}$ and the latter one as $\mathcal{M}^{CH}$ (see figure \ref{fig:SdSseparatecigar}). Furthermore, to make sure that in $\mathcal{M}^{BH}$ we do not include degrees of freedom near $\mathcal{M}^{CH}$ and vice versa, we define a thin shell with thickness $a$ which separates $\mathcal{M}^{BH}$ and $\mathcal{M}^{CH}$, where $a$ can be arbitrarily small. In this case, $\mathcal{M}^{BH}$ is defined for $r_{BH}<r<r_c-\frac{a}{2}$ while $\mathcal{M}^{CH}$ is defined for $r_c+\frac{a}{2}<r<r_{CH}$. We can then define codimension-1 slices $\Sigma^{BH}=\Sigma_A^{BH}\cup\Sigma_B^{BH}$ and $\Sigma^{CH}=\Sigma^{CH}_A\cup\Sigma^{CH}_B$ such that
\begin{align}
\Sigma_A^{BH}&=\{\tau,x^i|\tau=0,r\in[r_{BH},r_c-a/2]\},\\
\Sigma_B^{BH}&=\{\tau,x^i|\tau=\beta_{BH}/2,r\in[r_{BH},r_c-a/2]\},
\end{align}
and
\begin{align}
\Sigma_A^{CH}&=\{\tau,x^i|\tau=0,r\in[r_c+a/2,r_{CH}]\},\\
\Sigma_B^{CH}&=\{\tau,x^i|\tau=\beta_{CH}/2,r\in[r_c+a/2,r_{CH}]\}.
\end{align}
Here, $\beta_{BH}=\frac{4\pi}{f'(r_{BH})}$ and $\beta_{CH}=\frac{4\pi}{|f'(r_{CH})|}$ are the inverse temperature of the black hole horizon and the cosmological horizon respectively.\\
\indent The approximation $r_{BH}\ll r_c\ll r_{CH}$ is valid as long as the metric near $r_c$ can be approximated by Minkowski spacetime from both black hole and cosmological horizon perspective. Furthermore, we may add a boundary condition at $r_c$ such that the variation of the action at $r_c$ vanishes. To be precise, we impose the variation of the action at the boundaries located at both $r=r_c-a/2$ and $r=r_c+a/2$ vanish for regions near black hole horizon and cosmological horizon respectively. Therefore, the derivative of the action with respect to the R\'enyi index, $\partial_qI_E[\hat{\mathcal{M}}_q]$, is still given by eq. (\ref{derivativeaction}) for both black hole and cosmological horizon, and we may calculate the entanglement entropy in $\mathcal{M}^{BH}$ and $\mathcal{M}^{CH}$ from this. Since radius of the $\textbf{S}^2$ part of the near-horizon geometry for $\hat{\mathcal{M}}_q^{BH}$ and $\hat{\mathcal{M}}_q^{CH}$ is given by $r_{BH}$ and $r_{CH}$ respectively, the entanglement entropy is given by
\begin{equation}
S_A^{BH}=\pi r_{BH}^2,
\end{equation}
for the entanglement of region separated by the black hole and
\begin{equation}
S_A^{CH}=\pi r_{CH}^2,
\end{equation}
for the one that are separated by the cosmological horizon. Here, we have two distinct entanglement entropy describing entanglement in two disconnected regions $\mathcal{M}^{BH}$ and $\mathcal{M}^{CH}$.
 \\\\
 \textbf{Schwarzschild-de Sitter Without Near-Horizon Geometry}\\
We may extend the calculation of the metric away from the horizon in SdS case as well. Following similar reason with the RN case, here the only parameter which depends on the R\'enyi index $q$ is the mass, i.e. $M\rightarrow M_q$. This is so because we need to keep the asymptotic boundary condition where the metric is dominated by the $\frac{r^2}{L^2}$ part to be independent of the R\'enyi index $q$. Therefore, the $f(r)$ function for SdS is replaced by
\begin{equation}
f_q(r)=1-\frac{2M_q}{r}-\frac{r^2}{L^2},
\end{equation}
where $M_1=M$.\\
\indent Next, we consider the case for both black hole horizon and cosmological horizon. In the black hole region, we have (from eq. (\ref{fqconstraint}))
\begin{equation}
f'_q(r_{BH})=\frac{2M_q}{r_{BH}^2}-\frac{2r_{BH}}{L^2}=\frac{f'(r_{BH})}{q}.
\end{equation}
Therefore, the mass $M_q$ is given by
\begin{equation}
M_q=\frac{r_{BH}^2}{2}\bigg(\frac{f'(r_{BH})}{q}+\frac{2r_{BH}}{L^2}\bigg).\label{SdSMq}
\end{equation}
To calculate entanglement entropy via $\partial_qI_E[\hat{\mathcal{M}}_q]$, we need the derivative of $M_q$ with respect to $q$, which is given by
\begin{equation}
\frac{dM_q}{dq}=-\frac{r_{BH}^2f'(r_{BH})}{2q^2},
\end{equation}
and we have
\begin{equation}
\partial_q f_q(r)\big|_{q=1}=\frac{\partial f_q(r)}{\partial M_q}\frac{dM_q}{dq}\bigg|_{q=1}=\frac{r_{BH}^2f'(r_{BH})}{r},
\end{equation}
which gives $\partial_q f_q(r_{BH})\big|_{q=1}=r_{BH}f'(r_{BH})$. This again reproduces $\partial_q I_E[\hat{\mathcal{M}}_q]$ as in eq. (\ref{deractionresult}) with the radius replaced by $r_{BH}$ and gives us the area law entanglement entropy with black hole's area. For the cosmological horizon case, we have a similar result with the inner horizon of the RN black hole since the temperature is negative. Here, we have $\partial_q f_q(r_{CH})=-r_{CH}|f'(r_{CH})|$ and also reproduces the area law of entanglement entropy.\\
\indent 
\\\\
\textbf{Entanglement in "Lukewarm" case}\\
Now, we would like to consider the equilibrium case where $\beta_{BH}=\beta_{CH}$ can be achieved. To avoid the extreme case (Nariai solution), we work on the Reissner-Nordstr\"om-de Sitter (RNdS), which is given by eq. (\ref{staticsphericalbh}) with
\begin{equation}
f(r)=1-\frac{2M}{r}+\frac{Q^2}{r^2}-\frac{r^2}{L^2}.
\end{equation}
In this solution, we have three real and positive solutions to $f(r)=0$, which is given by the inner $r_-$, outer $r_+$ and the cosmological horizon $r_{CH}$. In general, i.e. when $\beta_{BH}\neq \beta_{CH}$, we may calculate the entanglement entropy of the inner horizon $r_-$ by the same method when we computing the one for the Reissner-Nordstr\"om black hole explained in the previous subsection. For the outer $r_+$ and cosmological horizon $r_{CH}$, we calculate the entanglement entropy by separating the cigar with a thin shell just like the normal SdS calculation which have been explained earlier.\\
\indent However, in a RNdS spacetime, we could have a solution with $\beta_{+}=\beta_{CH}$, without requiring the two horizons to be coincide. This is called the "lukewarm" case of the RNdS black hole \cite{PhysRevD.41.403}. This condition is achieved when $M=Q$ and the horizons are given by
\begin{align*}
r_-&=\frac{L}{2}\bigg(-1+\sqrt{1+\frac{4M}{L}}\bigg),\\
r_+&=\frac{L}{2}\bigg(1-\sqrt{1-\frac{4M}{L}}\bigg),\\
r_{CH}&=\frac{L}{2}\bigg(1+\sqrt{1-\frac{4M}{L}}\bigg).
\end{align*}
The temperature of the lukewarm black hole, $\beta_L^{-1}=\frac{f'(r_+)}{4\pi}=\frac{|f'(r_{CH})|}{4\pi}$, is given by
\begin{equation}
\beta_L^{-1}=T_L=\frac{1}{4\pi L}\sqrt{1-\frac{4M}{L}}.
\end{equation}
In this case, we have no problem with conical singularities in the region $r_+<r<r_{CH}$ since the periodicity required to remove conical singularity at $r=r_+$ and $r=r_{CH}$ is identical. Thus, again, the Euclidean cigar at the region $r_+<r<r_{CH}$ is depicted by figure \ref{fig:SdSLukewarmCigar}. We call this part of the cigar geometry $\mathcal{M}^L$, which stands for "lukewarm" case.\\
\indent We define a slice $\Sigma^L$ as usual as a slice which divides $\mathcal{M}^L$ into two time-symmetric sides. We then decompose $\Sigma^L$ into $\Sigma^L=\Sigma^L_A\cup\Sigma_B^L$ with $\partial\Sigma_A^L=\partial\Sigma_B^L$ be the entangling surfaces located at $r=r_{BH}$ and $r_{CH}$. The surfaces $\Sigma_A^L$ and $\Sigma_B^L$ can be written as
\begin{align}
\Sigma_A^L&=\{\tau,x^i|\tau=0,r\in[r_{BH},r_{CH}]\},\\
\Sigma_B^L&=\{\tau,x^i|\tau=\beta_L/2,r\in[r_{BH},r_{CH}]\}.
\end{align}
To calculate entanglement between $\Sigma_A^L$ and $\Sigma_B^L$, we need $\partial_q I_E[\hat{\mathcal{M}}_q^L]$, which is now given by
\begin{align}\label{variationactionlukewarm}
\partial_q I_E[\hat{\mathcal{M}}_q^L]=&\int_{\xi_+=\varepsilon}\frac{d^3x}{16\pi G_N}\sqrt{\gamma_{+}}n_\rho\big(g^{(q)}_{\mu\nu}\nabla^\rho\partial_q g^{(q)\mu\nu}-\nabla_\mu\partial_q g^{(q)\rho\mu}\big)\\&+\int_{|\xi_{CH}|=\varepsilon}\frac{d^3x}{16\pi G_N}\sqrt{\gamma_{CH}}n_\rho\big(g^{(q)}_{\mu\nu}\nabla^\rho\partial_q g^{(q)\mu\nu}-\nabla_\mu\partial_q g^{(q)\rho\mu}\big),\nonumber
\end{align}
where $\xi_+\equiv r-r_+$ and $|\xi_{CH}|=|r-r_{CH}|$ are the distances to the outer and cosmological horizon respectively. Here, $\gamma_+$ is the induced metric at the surface $\xi_+=\varepsilon$ and $\gamma_{CH}$ is the induced metric at the surface $|\xi_{CH}|=\varepsilon$.  The manifold $\hat{\mathcal{M}}_q^L$ have conical singularities at both $r=r_+$ and $r=r_{CH}$ with opening angle $\Delta\phi=2\pi(1-1/q)$, or in other words, we have two fixed points located at the two horizons. Therefore, there are two surface terms which contribute to $\partial_q I_E[\hat{\mathcal{M}}_q^L]$, which are given by the first and second term of eq. (\ref{variationactionlukewarm}).\\
\indent To calculate the entanglement entropy in lukewarm case, we need to know the metric of $\hat{\mathcal{M}}_q^L$ near the horizons $r_+$ and $r_{CH}$. Both near-horizon metrics for $r_+$ and $r_{CH}$ are given by
\begin{equation}
ds^2=\tilde{y}^2d\bigg(\frac{2\pi\tau}{\beta_L}\bigg)^2+d\tilde{y}^2+r_+^2d\Omega^2+...\;,
\end{equation}
and
\begin{equation}
ds^2=\tilde{y}^2d\bigg(\frac{2\pi\tau}{\beta_L}\bigg)^2+d\tilde{y}^2+r_{CH}^2d\Omega^2+...\;,
\end{equation}
respectively. Thus, the entanglement entropy of the lukewarm case is given by
\begin{equation}
S_A^L=\pi r_+^2+\pi r_{CH}^2.\label{lukewarmentropy}
\end{equation}
We obtain that in this case, the entanglement entropy is the sum of the outer horizon entropy $\pi r_+^2$ and the cosmological horizon entropy $\pi r_{CH}^2$. This entanglement entropy captures the correlations between degrees of freedom which are strong near the two horizons, $r_+$ and $r_{CH}$ as the leading term. \\
\indent Notice that the region away from $r_+$ does not contribute so much to the $\pi r_+^2$ part of the entanglement entropy; this is also true for the region away from $r_{CH}$. However, as we move quite far from $r_+$, we will eventually reach the region which is close enough to $r_{CH}$. In this region, the correlation near $r_{CH}$ becomes stronger. This is also true if we look from the cosmological horizon perspective, i.e. if we move far away from $r_{CH}$. Therefore, we suggest that there should be some correlation term to the entropy given by eq. (\ref{lukewarmentropy}), such that
\begin{equation}
S_A^L=\pi r_{CH}^2(1+x^2+g(x)),\label{Lukewarmentropy}
\end{equation}
where $x\equiv\frac{r_+}{r_{CH}}<1$ and $g(x)$ is some function of $x$ which represents correlation between two horizons. We cannot obtain the explicit value of the function $g(x)$ by using this method of calculating the entanglement entropy. However, we suggest that the function goes to zero as $x\rightarrow0$, meaning that the correlation vanishes when two horizons are separated far away. This condition is satisfied when $\frac{M}{L}\rightarrow0$, and the lukewarm RNdS approaches vacuum de Sitter limit. In small $x$, we expect that the function $g(x)$ only contributes as the correction term to the entropy given by eq. (\ref{lukewarmentropy}) recalling that the calculation of entanglement entropy by this method only captures the leading term. Therefore, for small $x$, the function should behaves as $g(x)\sim\mathcal{O}(x^3)$. \\
\indent In the Lukewarm case, we also have the inner horizon $r_-$, where the entanglement entropy calculations are similar to the inner horizon of the RN case and hence we are not discuss such case in detail.\\\\
\begin{figure}
\centering
\includegraphics[scale=0.9]{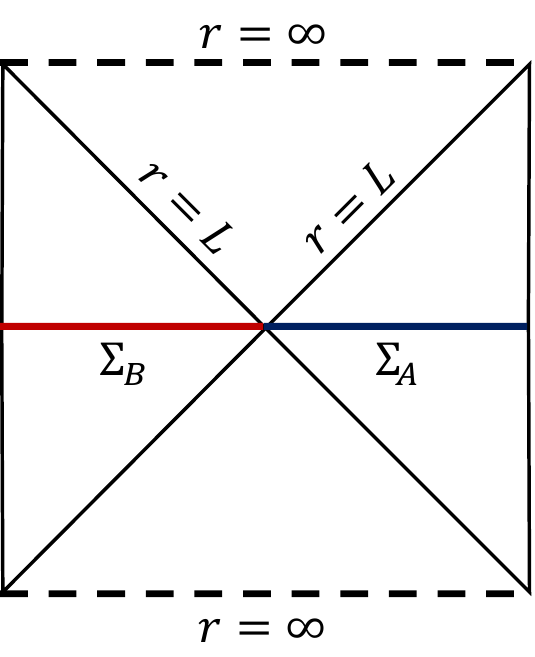}
\caption{The Carter-Penrose diagram of a de Sitter spacetime. The entanglement entropy calculates entanglement between regions $\Sigma_A$ and $\Sigma_B$, as a constant-time slice of a right and left de Sitter's Rindler wedge.}\label{fig:dSPenrose}
\end{figure}
\textbf{de Sitter limit ($M\rightarrow0$)}\\
Consider a vacuum de Sitter limit with $M\rightarrow0$, where the function $f(r)$ is now given by
\begin{equation}
f(r)=1-\frac{r^2}{L^2}.
\end{equation}
In Euclidean setting, we have $0<r<L$ where in this region, the value of $f(r)$ is positive. Therefore, the cigar geometry, in oppose of the previous examples (especially the Schwarzschild cigar), we have a finite-sized cigar which ends at $r=0$, where the geometry approaches Minkowski spacetime, and the tip is located at $r=L$. The de Sitter limit is recovered after taking the limit of $x\rightarrow0$ in eq.  (\ref{Lukewarmentropy}), or $M\rightarrow0$, which is nothing but the quarter of the cosmological horizon's area, $S_A=\pi L^2$, or the Gibbons-Hawking entropy \cite{GibbonsHawking_1977}. This entropy captures entanglement between regions on the right and the left Cauchy slice of de Sitter's Rindler wedge (see figure \ref{fig:dSPenrose}).\\
\indent The calculations of entanglement in de Sitter spacetime have been done earlier in \cite{Arias_2020}. In the aforementioned work, entanglement in the bulk de Sitter spacetime is given by twice of the area of the cosmological horizon divided by $4$, $S=2\pi L^2$, instead of just a single area, $S=\pi L^2$. The extra two factor arise because in that case, there are two fixed points corresponding to the fixed points located in the asymptotic boundary of the de Sitter spacetime, which then extends to the bulk spacetime. In this work, we consider a single fixed point in de Sitter bulk which divides two regions in the bulk located at the bifurcation point as can be seen in figure \ref{fig:dSPenrose}, which is not an extension of fixed points in the boundary to the bulk.\\
\indent The mass $M_q$ in de Sitter case also differs from the Schwarzschild-de Sitter case since now we require $M_1=0$. This is automatically satisfied in de Sitter spacetime because in the original spacetime, the value of $f'(L)$ for de Sitter spacetime is given by
\begin{equation}
f'(L)=-\frac{2}{L},
\end{equation}
and hence, by replacing $r_{BH}$ with $L$ in eq.  (\ref{SdSMq}), we have
\begin{equation}
M_q=L\bigg(1-\frac{1}{q}\bigg)=L\varepsilon_q,
\end{equation}
where we recover $M_1=0$. The derivative of the mass with respect to $q$ is given by
\begin{equation}
\frac{dM_q}{dq}=\frac{L}{q^2},
\end{equation}
and we have
\begin{equation}
\partial_qf_q(r)\big|_{q=1}=-\frac{2L}{rq^2},
\end{equation}
which then gives us the Gibbons-Hawking entropy $S=\pi L^2$. What interesting is that in the replicated manifold $\hat{\mathcal{M}}_q$ for de Sitter spacetime, we have an additional "mass" which is proportional to $\varepsilon_q$. Since we keep $\varepsilon_q$ to be small, we see that the replica trick calculation of empty de Sitter spacetime add a small amount of "mass" to the replica geometry.
\\\\
\textbf{Kiselev black hole}\\
Another black hole solution with cosmological horizon may arise from a solution called the Kiselev black hole \cite{Kiselev2003}, which is the solution of Einstein's equation with matter field obeying the equation of state
\begin{equation}
p_i=w_i\rho_i.\label{equationofstate}
\end{equation}
The metric of Kiselev black hole is given by the usual static and spherically symmetric black hole spacetime (\ref{staticsphericalbh}) with
\begin{equation}
f(r)=1-\frac{2M}{r}+\frac{Q^2}{r^2}-\sum_i\bigg(\frac{r_i}{r}\bigg)^{3w_i+1},
\end{equation}
where $r_i$ is some normalization constant with dimension length and $i$ denote the $i$-th matter satisfying the equation of state (\ref{equationofstate}). The term including the charge $Q$ arise if we consider a Reisnerr-Nordstr\"om-Kiselev black hole, which can simply be removed if we only care about the uncharged Kiselev black hole. For some quintessential matter (or dark energy which accelerates the universe), we have $-1<w_i<-1/3$.\\
\indent The Kiselev black hole with $-1<w_i<-1/3$ may give rise to a solution with cosmological horizon, similar to the de Sitter case. In fact, if we have $w=-1$, which corresponds to the cosmological constant, the solution reduces to the RNdS case. Another simple example is given by $w_i=-2/3$, where the function $f(r)$ now becomes
\begin{equation}
f(r)=1-\frac{2M}{r}+\frac{Q^2}{r^2}-\frac{r}{r_w}.
\end{equation}
From $f(r)=0$, we can have up to three roots which corresponds to the horizon. To classify the horizons as the inner, outer, and cosmological horizon, other than the sizes which usually obey $r_-<r_+<r_{CH}$, we could see in which region the function $f(r)$ is positive. For example, we could have a RNdS-like solution with $M=2, Q=1, r_w=20$ as can be seen in figure. In this case, the region $r_+<r<r_{CH}$ mimics the same region in the RNdS case. \\
\indent To calculate the entanglement entropy of a RN-Kiselev black hole, we may ask whether the value of $f'(r_+)$ and $f'(r_{CH})$ coincides or not, which corresponds to which method we should use, either the standard "cutting" procedure or the lukewarm case. The rest will give us a same result with the RNdS case, where the entanglement entropy of $r_+$ and $r_{CH}$ is separated if $f'(r_+)\neq f'(r_{CH})$ or gives an entanglement entropy in the form of eq. (\ref{lukewarmentropy}) if $f'(r_+)=f'(r_{CH})$. Thus, the information about the value of $f(r)$, i.e. where it becomes positive and where it becomes negative, is enough if we want to consider the entanglement entropy of the black hole spacetime. We then only need analogous interpretation to the RNdS case. We use this fact to study entanglement entropy in a more general black hole, such as black holes in the $F(R)$ theory of gravity, in the next section.
\subsection{Static and Spherically Symmetric Black Holes in $F(R)$ Gravity}\label{sec:3.3}
So far, we have calculated the entanglement entropy of various maximally-extended static and spherically black hole spacetimes in Einstein gravity.  In general, the entanglement entropy calculates the entanglement between two timelike regions separated by the horizon at the bifurcation surface which act as the entangling surface. In this section, we would like to consider a more general case where the Lagrangian is not given by Einstein theory such as in eq. (\ref{EinsteinHilbert}), but might consist of some higher-power of the curvature scalar. The gravitational action is now given by
\begin{equation}
I[\mathcal{M}]=\frac{1}{16\pi G_N}\int_{\mathcal{M}}d^Dx\sqrt{|G|}F(R),\label{FRaction}
\end{equation}
where $F(R)$\footnote{In many literature, the notation $f(R)$ is commonly used, instead of the capital $F(R)$, while $F(R)$ is usually used as the derivative $df(R)/dR$. However, in this work, we used the symbol $f(r)$ for the "blackening factor" of a black hole spacetime extensively. Hence, to avoid too much similarities, we use $F(R)$ instead and the derivative is symbolized by $F'(R)=dF(R)/dR$.} is some function of $R$, e.g. $F(R)=\sum_{n=1} \alpha_nR^n$, where $R$ is the usual Ricci scalar and $\alpha_n$ is some coefficients. In this case, we consider a higher-dimensional theory of gravity with dimension $D$ and $G$ is now a $D$-dimensional metric tensor. This is called the $F(R)$ gravity theory \cite{RevModPhys.82.451}. In general, the function $F$ might also depends on the Ricci tensor and Riemann tensor such that $F=F(R,R_{\mu\nu}R^{\mu\nu},R_{\mu\nu\alpha\beta}R^{\mu\nu\alpha\beta})$ but we keep it simple by considering only the $F(R)$ function.\\
\indent For practical purpose, we go straight to the variation of the (Euclidean) action of $F(R)$ gravity with respect to the R\'enyi parameter $q$, which is given by
\begin{align}\label{variationFR}
\partial_q I_E[\mathcal{M}]=&\int_{\mathcal{M}}\frac{d^Dx}{16\pi G_N}\sqrt{G}E_{AB}\partial_q G^{AB}\\\nonumber
&+\int_{\partial\mathcal{M}}\frac{d^{D-1}x}{16\pi G_N}\sqrt{\Gamma}\hat{n}_CF'(R)\big(G_{AB}\nabla^C\partial_q G^{AB}-\nabla_D\partial_q G^{CD}\big)\\\nonumber
&+\int_{\partial\mathcal{M}}\frac{d^{D-1}x}{16\pi G_N}\sqrt{\Gamma}\nabla_DF'(R)\big(\partial_q G^{CD}-G^{CD}G_{AB}\partial_q G^{AB}\big).
\end{align}
Here, $F'(R)\equiv\frac{dF(R)}{dR}$ denotes the derivative of the function $F(R)$ with respect to the Ricci scalar $R$, $\Gamma$ is a $(D-1)$-dimensional induced metric in the boundary $\partial\mathcal{M}$, $\hat{n}^C$ is the unit normal vector of the boundary $\partial\mathcal{M}$, and $E_{AB}=0$ is the equation of motion of the $F(R)$ theory. Note that in $F(R)$ gravity, we have two surface terms, where the third term of eq.  (\ref{variationFR}) is equal to zero in Einstein gravity since in this case, $F'(R)=1$ and thus the derivative vanishes. \\
\indent When solving the equation of motion in $F(R)$ gravity, we might expect black hole solutions in the form of (see, e.g. \cite{Tang2021})
\begin{equation}
ds^2=-f(r)dt^2+\frac{dr^2}{f(r)}+r^2\Omega_{ij}(x)dx^i dx^j,
\end{equation}
where $i,j$ runs from $2$ to $D$. The black hole horizons are located, as usual, at the solutions of $f(r)=0$. Now consider a black hole solution with a single horizon $r_H$. Following previous steps in calculating the entanglement entropy, we expand the metric near the horizon and perform some redefinition of the coordinates to obtain a metric in the form of
\begin{equation}
ds^2=\frac{\tilde{y}^2}{q^2}d\tilde{\tau}^2+d\tilde{y}^2+r_H^2\Omega_{ij}(x)dx^idx^j+\mathcal{O}(\tilde{y}^2),\label{nearhorizonfr}
\end{equation}
where $\tilde{y}$ is again proportional to the (square root of the) distance to the horizon and $\tilde{\tau}$ is proportional to the Euclidean time, which is periodic.\\
\indent We then plug the near horizon metric of eq. (\ref{nearhorizonfr}) into the derivative of the action (\ref{variationFR}) and integrate $\tilde{\tau}$ from $0$ to $2\pi$, which gives
\begin{align}\label{frderactionresult}
\partial_qI_E[\hat{\mathcal{M}}_q]&=\frac{1}{16\pi G_N}\times 2\pi\times\frac{\tilde{y}}{q}\times\frac{2}{\tilde{y}q}\times\int d^{D-2}xr_H^2\sqrt{\Omega}\times F'(\bar{R})+\mathcal{O}(\tilde{y})\bigg|_{\tilde{y}=0}\\\nonumber
&\;\;\;\;\;\;+\frac{1}{16\pi G_N}\int d^{D-1}x\nabla_{\tilde{\tau}} F'(\bar{R})\times\frac{\tilde{y}}{q}\times\frac{2}{q}\times r_H^2\sqrt{\Omega}+\mathcal{O}(\tilde{y}^2)\bigg|_{\tilde{y}=0}\\\nonumber
&=\frac{\text{Area(Horizon)}}{4G_Nq^2}F'(\bar{R})\bigg|_{\tilde{y}=0},
\end{align}
where $\bar{R}$ is the Ricci scalar for the metric which is the solution to the equation of motion and 
\begin{equation}
\text{Area(Horizon)}=\int d^{D-2}xr_H^2\sqrt{\Omega},
\end{equation}
is the area of the black hole horizon with radius $r_H$ where $\Omega$ is the determinant $\det(\Omega_{ij})$. Notice that the second term of the first equation in (\ref{frderactionresult}) is zero in the limit $\tilde{y}=0$ and hence the entanglement entropy is again proportional to the area, or 
\begin{equation}
S_A=\frac{\text{Area(Horizon)}}{4G_N}F'(\bar{R})\bigg|_{r=r_H},
\end{equation}
with an extra factor $F'(\bar{R})$ evaluated at the horizon. This reproduces the calculation of the entanglement entropy for a static and spherically-symmetric black hole solution in $F(R)$ gravity, which has been studied earlier \cite{zheng_horizon_2018,PhysRevD.99.104018}.
\section{Conclusions and Discussions}\label{sec:4}
In this work, we study how to calculate entanglement entropy in (maximally-extended) black hole spacetimes, for two regions separated by the horizon (wormhole). The replica trick is used in calculating the entanglement entropy, i.e. we calculate the $q\rightarrow1$ limit of the R\'enyi entropy. Calculating the limit means we evaluate $\partial_q I_E[\hat{\mathcal{M}}_q]$ at $q=1$ and hence we involve the calculation of the boundary term of the gravitational action for the replica manifold $\hat{\mathcal{M}}_q$. We consider first the Schwarzschild case, where there is only one horizon separating two regions. Furthermore, we extend the calculation to include more general static and spherically-symmetric black holes which may have several horizons. Some specific cases are considered, for example, the Reissner-Nordstr\"om and Schwarzschild-de Sitter black holes which represent black holes with an inner horizon and cosmological horizon respectively.\\
\indent In this section, we sum everything up and elaborate on some step-by-step ways to calculate entanglement entropy in black hole spacetimes:
\begin{enumerate}
\item We need to have a static and spherically-symmetric non-extreme black hole spacetime $\mathcal{M}$ in the form of eq. (\ref{staticsphericalbh}) which is the solution of the gravitational equation of motion, and consider how many horizons are available in the solution.  Then, define regions with $f(r)>0$ and their corresponding horizon. This helps us to consider whether our case is represented by the RN or S-dS black hole (or maybe both), as explained in Sections 3.1 and 3.2.
\item Construct the replica manifold $\hat{\mathcal{M}}_q$ which corresponds to the original black hole manifold $\mathcal{M}$. The metric $ds^2$ for $\hat{\mathcal{M}}_q$ can be obtained, as explained in this work when calculating the general static and spherically-symmetric case without near-horizon limit at the beginning of Section 3, by considering some requirements:
\begin{enumerate}
\item The replica manifold $\hat{\mathcal{M}}_q$ has a conical singularity located at the horizon (singularities, when there is more than one horizon) with deficit angle $\Delta\phi=2\pi(1-1/q)$. This gives us the near-horizon solution to the metric in the form of eq. (\ref{nearhorizongeneral}), which is smooth where $\tilde{\tau}\sim\tilde{\tau}+2\pi q$, but we keep the periodicity of $\tilde{\tau}$ to be $2\pi$.
\item The replica manifold $\hat{\mathcal{M}}_q$ satisfies the gravitational equation of motion at least at the first order of $\varepsilon_q=1-1/q$. Following this and the requirement in (a), this gives us the constraint in the eq. (\ref{fqconstraint}), stating that we may only work for small value of $\varepsilon_q$ (only up to the first order of $\varepsilon_q$) and the original function $f_q(r)$ can be determined by its first derivative via $f_q'(r_H)=f'(r_H)/q$.
\end{enumerate}
\item Find which black hole parameter(s)---such as mass, charge, spin---that may become the function of $q$ by keeping the desired boundary conditions. For example, in RN black hole, the charge $Q$ cannot depend on the R\'enyi index $q$ due to the boundary condition near the true singularity. Then, find the explicit value of the parameter(s) as the function of $q$, plug this back into the function $f_q(r)$ to obtain the metric for $\hat{\mathcal{M}}_q$. Note that we can only change the black hole parameter to become the function of $q$ since, in this way, we do not change the geometry. For example, Schwarzschild black hole remains a Schwarzschild black hole in $\hat{\mathcal{M}}_q$ but with mass replaced by $M\rightarrow M_q=M/q$.
\item To calculate the entanglement entropy, we use the obtained information (the explicit form of $f_q(r)$), and plug it into the eq. (\ref{deractiongeneral}). Multiply the result by $q$ and take the limit $q\rightarrow1$ gives us the entanglement entropy. In every case that we consider in this work, all entanglement entropies lead to the usual area law and hence reproduce the Bekenstein-Hawking entropy.
\end{enumerate}
We find that the conclusion that the entanglement entropy reproduces the Bekenstein-Hawking area law applies to solutions with matter action and $F(R)$ gravity as well.\\
\indent For black holes with inner horizon such as the charged RN black hole, we obtain that there are two distinct entanglement entropies. The first one corresponds to the regions separated by the outer horizon and the other one corresponds to the regions separated by the inner horizon, excluding the true singularity located at $r=0$. In the extreme limit, the entanglement entropy approach the Bekenstein-Hawking entropy for an extreme black hole with the extreme degenerate horizon radius $r_H$. However, at the extreme point itself, the entanglement entropy is zero since there is no region in the maximally-extended spacetime which can be separated by the horizon. Hence, in the extreme case, the density matrix of the region under consideration is pure and the entanglement entropy becomes zero, indicating the possibility of a connection with phase transition \cite{Pavan1991}.\\
\indent For black holes with a cosmological horizon, the entanglement entropy is given by the sum of the entropy of the black hole and the cosmological horizon, if the black hole and cosmological horizon are separated quite far away. Otherwise, we need to consider the lukewarm case where the temperature of the black hole and the cosmological horizon are identical, to avoid conical singularity in the original metric $\mathcal{M}$ which cannot be removed by defining the periodicity of the Euclidean time. Since we can only fix one periodicity for the Euclidean time, if the temperatures are different, there exists a conical singularity located at the horizon in which the temperature does not match the periodicity of the Euclidean time. In a lukewarm case, we may have some additional function to the entropy which represents the correlation between a black hole and cosmological horizon entropy which is of order $\mathcal{O}(x^3)$ for small $x\equiv \frac{r_{BH}}{r_{CH}}$. However, we cannot obtain the explicit form of the function and hence we keep the discussion to be qualitative. Furthermore, entanglement entropy for the extreme Nariai limit does not exist since there is no timelike region/slice to be considered.\\
\indent For more general black holes with both inner and cosmological horizons, the generalization is rather straightforward. The regions which correspond to the RN-like case and the one with the S-dS-like case are separable. Hence, we expect that the entanglement entropy consists of the one for the inner horizon and the one for the outer and cosmological horizons. If there exist degenerate horizons, indicating extreme cases at the corresponding horizons, then the entanglement entropy becomes zero if the region is RN-like. However, we do not consider the extreme case for the S-dS-like region since the notion of entanglement entropy cannot be defined there.\\
\indent Since the general calculation without near-horizon geometry states that we can only work for $q$ near 1, we may not be able to calculate the entropy with arbitrary $q$ such as the R\'enyi or modular entropy. Calculations involving higher order of $\varepsilon_q$ will be presented elsewhere. However, since the entanglement entropy is only the $q\rightarrow1$ limit of either the R\'enyi or modular entropy, there is no problem in calculating the entanglement entropy. Calculations of one-parameter generalization to the entanglement entropy (R\'enyi or modular entropy) for arbitrary $q$ remain unknown. \\
\indent In addition, it might also be interesting to consider entanglement in rotating black hole spacetimes. However, this might add some difficulties in calculating the entropy via replica trick since the metric is now stationary with crossing terms between spatial and time coordinates. Thermodynamics in rotating black holes has been studied many times, but the study of their entanglement properties is still lacking. Since rotating black holes also have a similar Carter-Penrose diagram with the Reissner-Nordstr\"om case, we might also expect similar results. Furthermore, calculating entanglement entropy of general rotating black holes from their symmetry (in this case, axial symmetry) is also interesting. The study of entanglement entropy which focuses on rotating black holes will be presented elsewhere. 
\section*{Appendix A: Near Horizon Geometry from Gaussian Normal Coordinates}
Here we expand the near-horizon geometry using the Gaussian Normal Coordinates to the next subleading order of $\tilde{y}$:
\begin{equation}
ds^2=q^2\bigg(\frac{\tilde{y}^2}{q^2}d\tilde{\tau}^2+d\tilde{y}^2\bigg)+(\gamma_{ij}+2K_{ij}^1\tilde{y}^q\cos\tilde{\tau}+2K_{ij}^2\tilde{y}^q\sin\tilde{\tau})dy^idy^j+...\;,
\end{equation}
where $K_{ij}^{1,2}$ are the extrinsic curvature with respect to two vectors $n_\mu^{1,2}$ which are normal to the hypersurface $r=r_H$ and $\tau=const.$. The metric is obtained by expanding the total metric of the manifold $\mathcal{M}$ near the horizon hypersurface and requiring the time coordinate $\tilde{\tau}$ to be periodic with periodicity $\tilde{\tau}\sim\tilde{\tau}+2\pi$ and have a conical singularity located at $\tilde{y}=0$ with deficit angle $\Delta\phi=2\pi(1-1/q)$.\\
\indent Plugging this into the derivative of the action $\partial_qI_E[\hat{\mathcal{M}}_q]$ such as in eq. (\ref{derivativeaction}) gives us the term which is proportional to $\tilde{y}\partial_q\tilde{y}^q\big|_{\tilde{y}=0}=\tilde{y}^{q+1}\log\tilde{y}\big|_{\tilde{y}=0}$, which is zero in the subleading term. Thus, even though we keep the next subleading order of $\tilde{y}$ in the near-horizon metric, the solution is still given by the area law, which confirms our claim made in Sec. 2.2.
\section*{Appendix B: Calculations With the Original $\hat{\mathcal{M}}_q$ Metric}
The full metric of the Schwarzschild solution for arbitrary $q$ is given by
\begin{equation}
ds^2=\bigg(1-\frac{r_S}{rq}\bigg)d\tau^2+\frac{dr^2}{(1-r_S/rq)}+r^2d\Omega^2,
\end{equation}
where $r_S=2M$ is the Schwarzschild radius. If we directly plug this metric into the calculation of $\partial_qI_E[\hat{\mathcal{M}}_q]$ and integrating $\tau$ from $0$ to $\beta=8\pi M$, we will have
\begin{align}
\partial_qI_E[\hat{\mathcal{M}}_q]&=\int_{\partial\hat{\mathcal{M}}_q(r=2M+\epsilon)}\frac{d^3x}{16\pi G_N}\sqrt{\gamma}\hat{n}_\rho(g_{\mu\nu}\nabla^\rho\partial_qg^{\mu\nu}-\nabla_\mu\partial_qg^{\mu\rho})\\\nonumber
&=\frac{1}{16\pi G_N}\times (8\pi M)\times(4\pi)\times\frac{2M}{q^2}\times\frac{3q+1}{q-1},
\end{align}
which is divergent in the limit $q\rightarrow1$.
\section*{Acknowledgement}
FPZ would like to thank Kemenristek, the Ministry of Research, Technology, and Higher Education, Republic of Indonesia for financial support. HLP would like to thank GTA Institut Teknologi Bandung for financial support. HLP would like to thank the members of the Theoretical Physics Groups of Institut Teknologi Bandung for their hospitality.
\section*{Data Availability}
Data sharing not applicable to this article as no datasets were generated or analysed during the current study.
\bibliographystyle{elsarticle-num}
\bibliography{EEBHSRT.bib}
\end{document}